\documentclass[12pt, a4paper]{article}
\pdfoutput=1

\usepackage{hyperref} 

\usepackage{float}  
\usepackage{easybmat}  
\usepackage{amsmath}
\usepackage{amsfonts}
\usepackage{multirow}
\usepackage{makecell}
\usepackage{amssymb}
\usepackage{graphicx, rotating}
\usepackage{epstopdf}
\usepackage{epsfig}
\usepackage{latexsym}
\usepackage{graphicx}
\usepackage{color}
\usepackage{bm,amssymb}
\usepackage{cite}
\usepackage{slashed}
\usepackage{arydshln}
\hypersetup{colorlinks, citecolor=bluscuro, linkcolor=black, urlcolor=bluscuro}
\definecolor{rossos}{cmyk}{0,1,1,0.55}
\definecolor{bluscuro}{rgb}{0.15, 0.2, .85}
\definecolor{bluchiaro}{cmyk}{1,.3,0.,0.1}

\usepackage[usenames,dvipsnames]{xcolor}
\usepackage{array}
\newcolumntype{C}[1]{>{\centering\let\newline\\\arraybackslash\hspace{0pt}}m{#1}}

\usepackage{enumerate}

\usepackage{comment}

\hypersetup{colorlinks, bookmarksnumbered, citecolor=darkblue, linkcolor=darkblue, pdfstartview=FitH, urlcolor=darkblue, linktocpage}

\definecolor{darkblue}{cmyk}{1,0.5,0,0.2}

\newcommand{\arXhref}[1]{\href{http://arxiv.org/abs/arXiv:#1}{#1}} 

\makeatletter

\@addtoreset{equation}{section}

\makeatother

\newcommand{\eq}[1]{Eq.~(\ref{#1})}
\newcommand{\lag}{\mathcal{L}}

\newcommand{\Dslash}{\!\not\!\!  D}
\newcommand{\be}{\begin{equation}}
\newcommand{\ee}{\end{equation}}
\newcommand{\bea}{\begin{eqnarray}}
\newcommand{\eea}{\end{eqnarray}}

\newcommand{\Op}{\mathcal{O}}

\def\lra#1{\overset{\text{\scriptsize$\leftrightarrow$}}{#1}}

\begin{document}

\thispagestyle{empty}

\begin{center}

\hfill SISSA  57/2013/FISI \\

\begin{center}

\vspace*{0.5cm}

{\Large\bf  Scaling and tuning of EW and Higgs observables}

\end{center}

\vspace{1.4cm}

{\bf Joan Elias-Mir\'o$^{a,b}$, Christophe Grojean$^{b,c}$, Rick S. Gupta$^{b}$, David Marzocca$^{d}$}

\vspace{1.2cm}

\vspace{.3cm}

${}^a\!\!$ {\em Dept. de F\'isica, Universitat Aut\`onoma de Barcelona, 08193 Bellaterra, Barcelona.}

${}^b\!\!$
{\em IFAE, Universitat Aut\`onoma de Barcelona, E-08193 Bellaterra, Spain}

${}^c\!\!$
{\em ICREA, 08010 Barcelona, Spain}

${}^d\!\!$
{\em SISSA and INFN, Via Bonomea 265, I-34136 Trieste, Italy}

\end{center}

\vspace{0.8cm}

\centerline{\bf Abstract}
\vspace{2 mm}
\begin{quote}

We study deformations of the SM via higher dimensional operators.  
In particular, we explicitly calculate the one-loop anomalous dimension matrix for 13 bosonic dimension-6 operators
relevant for electroweak and Higgs physics. 
These scaling equations allow us to derive RG-induced bounds, stronger than the direct constraints,
on a universal shift of the Higgs couplings and some anomalous triple gauge couplings by assuming no tuning at the scale of new physics, i.e. by requiring that their individual contributions to the running of other severely constrained observables, like the electroweak oblique parameters or $\Gamma(h \rightarrow \gamma\gamma)$, do not exceed their experimental direct bounds.
We also study operators involving the Higgs and gluon fields. 
\end{quote}

\vfill

\newpage

\tableofcontents

\section{Introduction}

The first run of the LHC, with the discovery of the Higgs boson and the measurement of its characteristic properties~\cite{Higgs1,Higgs2}, has been extremely successful. There are, however, still no signs of any new physics that can stabilize the Higgs mass and thus a deeper understanding of the electroweak (EW) symmetry breaking sector of the Standard Model (SM) is still missing.  All natural explanations for the EW symmetry breaking introduce new particles around the TeV scale which, when integrated out, alter the Higgs properties. Hence, measuring the Higgs sector with great accuracy has the potential of  clarifying the origin of EW symmetry breaking. Assuming  a mass gap between the SM scale and the new physics scale, as the lack of evidence for new physics seems to suggest, the Higgs properties and its deviations from the SM can be conveniently parametrized and systematically studied by higher dimensional operators~\cite{Buchmuller:1985jz},
\be
\delta {\cal L}= \sum\limits_{i} \frac{c_i}{\Lambda^2} {\cal O}_i \ ,
\ee
where the $c_i$'s are dubbed Wilson coefficients.

While these higher dimensional operators are generated at the new physics scale $\Lambda$, they are measured at  the lower scale of the experiments.\footnote{We assume that, at the scale $\Lambda$, the baryon and lepton numbers are conserved.}
 Due to renormalization group (RG) flow, the Wilson coefficients run and mix as we go down from $\Lambda$ to the experimental scale $\sim m_W$. The  operator coefficients at the two different scales are related to each other via the so-called anomalous dimension matrix. These quantum corrections mix the operators among themselves and therefore open the possibility of linking different kinds of deformations from the SM which are otherwise unrelated. In this article, we compute the anomalous dimension matrix of   a set of 13  dimension-6 (dim-6) operators  composed only of gauge bosons and Higgs fields and estimate the impact of these RG mixing effects  on experimental measurements. Some elements of the anomalous dimension matrix have been previously calculated in the  literature, see refs.~\cite{Narison:1983,Morozov:1985ef,Hagiwara:1993ck,Hagiwara:1994pw,Alam:1997nk,Barbieri:2007bh,Grojean:2013kd,Elias-Miro:2013gya,EEMP2,Jenkins:2013zja,Jenkins:2013wua}, with a  renewed interest after the recent Higgs property measurements.

To be completely general about the possible new physics scenarios one would need  to compute the anomalous dimension matrix for all the 59 dim-6 operators~\cite{Buchmuller:1985jz,Grzadkowski:2010es}.\footnote{This is the number of independent operators for one generation of fermions, see next section.} A given set of experimental observables, however,  receives contributions  only from a subset of these operators. The  dim-6 operators we are focussing our attention on, is a particularly interesting subset as they capture most of the possible deformations of the electroweak sector studied at LEP (i.e., electroweak precision tests and triple gauge couplings) and of the Higgs sector being currently studied at the LHC. At the same time, these operators are among the most important ones generated by universal new physics theories.\footnote{By universal theories we mean theories in which the BSM sector is flavour universal and in addition any new vector state couples to fermions via the SM SU(2)$\times$U(1) currents. } 
See for instance refs. \cite{Contino:2013kra, Dumont:2013wma} and refs. \cite{EEMP2,Pomarol:2013zra} for a recent general phenomenological analysis of the SM operators; the last two stress  the presence of blind directions on certain combinations of the Wilson coefficients\, \cite{DeRujula:1991se}. 

One may naively think that these  RG effects do not have a significant impact on phenomenology since they are loop suppressed. This is, however, not the case because the different Wilson coefficients have been constrained at different levels of precision. In particular, the ones contributing to LEP electroweak precision observables have been measured at the per mille level, whereas those parametrizing triple gauge couplings (TGC) and Higgs coupling data have been  measured at most at the percent level. This hierarchy in the size of constraints means that, despite the one loop factor, the  RG contributions of a weakly constrained coupling to a strongly constrained one can  be of the same order as, or even larger than, the bound on the strongly constrained coefficient. This means that the RG-mixing effects of such weakly constrained Wilson coefficients can be measured/constrained by precision measurements of other couplings to which experiments are more sensitive. Indeed, we find interesting instances of coefficients which receive stronger bounds from the RG mixing than from the direct tree-level constraint.  For example, we show that the Wilson coefficients parametrizing  deviations in some of the anomalous TGC observables and the correction to the Higgs kinetic term $\hat{c}_H$  receive a  stronger bound via their RG-mixing contribution to the electroweak parameters $\hat{S}$, $\hat{T}, W$,  $Y$ and $\Gamma_{h \rightarrow \gamma\gamma}$ than the direct constraint. In refs. \cite{Hagiwara:1993ck,Alam:1997nk}, and more recently refs. \cite{Chen:2013kfa,Mebane:2013zga}, the RG effects of the mixing of TGC and the EW parameters are studied.

The paper is organized as follows. In Section~\ref{Dim6Basis} we define the basis of dim-6 operators we shall use. Then, in Section~\ref{oneLoopRGES} we present our result for the anomalous dimension matrix of the 10 bosonic\footnote{By \emph{bosonic} operators we denote those operators made out of boson fields.} operators  related to EW and Higgs observables. In Section~\ref{Sconst} we shall use the RG equations (RGE's) to set bounds on the value of some Wilson coefficients that are otherwise less constrained by direct measurements;  we also comment on the future prospects.
In Section~\ref{GluonSect} we present the anomalous dimension matrix for a set of operators with gluons and discuss the available bounds on them.
We conclude in Section~\ref{Sec:Conclusions}.
 In the appendices, we report several details of our computations and present a comparison with existing results in literature.

\section{The dimension-six operator basis}
\label{Dim6Basis}

In this section we  define our choice for the dim-6 operator basis~\cite{Buchmuller:1985jz,Grzadkowski:2010es} and the subset of the dim-6 operators for which we want to compute the anomalous dimension submatrix.  Our choice of basis is motivated by the observables we are interested in, and the subset we consider is defined by the operators in this basis which give a tree-level contribution to our set of observables. In this work, we shall be interested in  EW observables, Higgs couplings to gauge bosons and QCD observables involving  gluons only and the relations among each other as imposed from the running between the scale of new physics to the weak scale. These include the four electroweak oblique pseudo-observables $\hat S$, $\hat T$, $W$ and $Y$, the three triple gauge coupling observables $ g_1^Z, \kappa_\gamma$  and $\lambda_\gamma$, the Higgs couplings to vector bosons, the gluon oblique parameter $Z$~\cite{Barbieri} and the anomalous triple gluon coupling parameter $\hat{c}_{3G}$. We describe these observables in more detail in Section~\ref{constraints} and Section~\ref{GluonSect}. 
For ealier systematic studies of the effects of higher-dimensional operators on these observables, see refs.~\cite{Han:2004az,Cacciapaglia:2006pk}.

We have not included the Higgs decays to fermions in our list of observables. The only dim-6 operators contributing to these observables are the operators  ${\cal O}_{y_u},  {\cal O}_{y_d}$ and ${\cal O}_{y_e}$, defined in ref.~\cite{EEMP2}.\footnote{The flat direction~\cite{Grojean:2013nya} between the operators ${\cal O}_{y_u}, {\cal O}_{BB}$ and ${\cal O}_{GG}$ from the measurements of Higgs couplings to photons and gluons is lifted by considering the (still loose) upper limit on the cross section production of a Higgs boson in association with a pair of top-antitop quarks~\cite{Pomarol:2013zra}. Stronger bounds on the Wilson coefficients of ${\cal O}_{BB}$ and ${\cal O}_{GG}$ can be obtained by imposing some theoretical priors on the value of the Wilson coefficient of  ${\cal O}_{y_u}$ but we did not consider these stronger bounds here and we can safely ignore the operator ${\cal O}_{y_u}$ in our analysis.} The RG effects  of these operators have been already studied in ref.~\cite{EEMP2}. These are  weakly constrained operators and  new RG-induced constraints  can be derived only if they contribute to the running of more strongly constrained operators. In ref.~\cite{EEMP2} it has been shown that there is no such contribution and therefore we do not include these operators in our analysis.

Before defining our choice for the dim-6 operator basis, let us specify the subset of independent operators on which we concentrate and which are part of the basis. This subset, which has the property that it can efficiently parametrize dim-6  contributions to the observables specified above, is given in Table~\ref{table:one1}. The basis therefore contains a total of 14 CP-even bosonic operators, notice however that ${\cal O}_6$ does not contribute to any of the observables we are interested in, neither at tree-level nor by RG running~\cite{EEMP2}; it contributes instead to the Higgs self-coupling which however is still not directly measured. For this reason  we did not include this observable in our list and did not compute its RG scaling.

The conventions in Table~\ref{table:one1} and in the rest of the text are as follows. We define $D_{\rho} W^a_{\mu \nu} = \partial_\rho W^a_{\mu \nu} + g \epsilon^{abc} W^b_\rho W^c_{\mu\nu}$, $H^\dagger {\lra { D_\mu}} H\equiv H^\dagger D_\mu H - (D_\mu H)^\dagger H $, with $D_\mu H = \partial_\mu H -i g\tau^a W^a_\mu H - i g' Y_H B_\mu H$. We have taken the hypercharge of the Higgs $Y_H=1/2$ and  $\tau^a=\sigma^a/2$ are the $SU(2)_L$ generators in the fundamental representation.

{\renewcommand{\arraystretch}{1.3} 
\begin{table}[tc]
\small
\centering
\begin{tabular}{ccc}
\begin{tabular}{|c|}\hline
${\cal O}_H=\frac{1}{2}(\partial^\mu |H|^2)^2$\\
${\cal O}_T=\frac{1}{2}\left (H^\dagger {\lra{D}_\mu} H\right)^2$\\
${\cal O}_6=\lambda |H|^6$ \\
${\cal O}_W=ig\left( H^\dagger  \tau^a \lra {D^\mu} H \right )D^\nu  W_{\mu \nu}^a$\\
${\cal O}_B=ig'  Y_H \left( H^\dagger  \lra {D^\mu} H \right )\partial^\nu  B_{\mu \nu}$\\
${\cal O}_{2W}=-\frac{1}{2}  ( D^\mu  W_{\mu \nu}^a)^2$\\
${\cal O}_{2B}=-\frac{1}{2}( \partial^\mu  B_{\mu \nu})^2$\\
${\cal O}_{2G}=-\frac{1}{2}  ( D^\mu  G_{\mu \nu}^A)^2$ \\ \hline 
\end{tabular}
&\qquad \qquad&
\begin{tabular}{|c|}\hline
${\cal O}_{BB}=g^{\prime 2} |H|^2 B_{\mu\nu}B^{\mu\nu}$\\
${\cal O}_{WB}= g{g}^{\prime} H^\dagger \sigma^a H W^a_{\mu\nu}B^{\mu\nu}$\\
${\cal O}_{WW}=g^2 |H|^2 W^a_{\mu\nu}W^{a \mu\nu}$\\
${\cal O}_{GG}=g_s^2 |H|^2 G_{\mu\nu}^A G^{A\mu\nu}$\\
${\cal O}_{3W}= \frac{1}{3!} g\epsilon_{abc}W^{a\, \nu}_{\mu}W^{b}_{\nu\rho}W^{c\, \rho\mu}$\\
${\cal O}_{3G}= \frac{1}{3!} g_s f_{ABC}G^{A\, \nu}_{\mu}G^{B}_{\nu\rho}G^{C\, \rho\mu}$
 \\\hline
 \end{tabular}
\end{tabular}
\caption{\small The 14 CP-even operators made of SM bosons. 
The operators have been grouped in two different categories corresponding to operators of the form $\text{(SM current)}\times\text{(SM current)}$ (left box) and operators which are not products of SM currents (right box). \label{table:one1}}
\end{table}}

Note that the four precision parameters $\hat S$, $\hat T$, $W$ and $Y$, generated in our basis by four bosonic dim-6 operators \cite{Grinstein:1991cd,Barbieri}, as we show in Section~\ref{constraints}, are sufficient to describe all possible dim-6 contributions to the $e^+e^-\rightarrow f^+f^-$  observables at  LEP 1 and 2,  only in the limit of universal new physics. To be completely general about possible new physics scenarios it would be necessary to include two more operators that contribute to the $e^+e^-\rightarrow f^+f^-$ experiment~\cite{EEMP2,Pomarol:2013zra},
\begin{equation}
{\cal O}_L=   (i H^\dagger \lra{D}_\mu H)(\bar{L}_L  \gamma^\mu L_L) \ ,  ~~~{\cal O}_{LL}^{1,2}=(\bar{L^1}_L \sigma^a \gamma^\mu L^1_L)(\bar{L}^2_L \sigma^a \gamma^\mu L^2_L) \ ,
\end{equation}
where the former affects the SM coupling of the $Z$ boson to the left-handed leptons, and the latter affects the measurement of $G_F$ (recall that the super-indices denote the fermion family). There are enough measurements to simultaneously constrain all six operators at the per mille level~\cite{leptgc}. The RG contributions of $\{{\cal O}_L, {\cal O}_{LL}^{1,2}\}$ to the other operators have been already computed and can be found in ref.~\cite{EEMP2}. We have not studied possible RG-contributions of the operators of Table~\ref{table:one1} to $\{{\cal O}_L, {\cal O}_{LL}^{1,2}\}$, such RG-contributions could be used to impose some bounds on the weakly constrained operators of Table~\ref{table:one1}, since  $\{{\cal O}_L, {\cal O}_{LL}^{1,2}\}$,  are constrained at the permil level~\cite{Pomarol:2013zra}. Such an analysis would require computing many more elements of the full anomalous dimension matrix as well as enlarging the list of observables under consideration; this analysis would be interesting but beyond the scope of the present project.

The operators in Table~\ref{table:one1} have been grouped in two different categories, corresponding to operators of the form $\text{\emph{(SM current)}}\times\text{\emph{(SM current)}}$ (left  box) and operators which are not products of SM currents (right box). There are also  6 CP-odd counterparts of the operators in the second box which complete the list of bosonic operators of our basis, 20 in total. The current-current operators can be related to each other and to other fermionic current-current operators, using the SM equations of motion (EoM) or, equivalently, by performing field redefinitions. As we discuss in the following paragraphs, this means that we have to be careful in choosing the other operators in our basis to ensure that there are no redundancies. As we discuss in Appendix~\ref{appenB}, these relationships give us an important consistency check on the anomalous dimension matrix we obtain. 

Although we are interested in the anomalous dimension matrix of the 13 operators in Table~\ref{table:one1}, we have to define the complete basis of dim-6 operators that we are using. This is because, as we shall see, under RG scaling many redundant operators  not in our basis, including operators containing fermions,  can be generated radiatively. These operators then need to be redefined away in terms of the ones in our basis. To clearly identify these redundant operators it is thus necessary to unambiguously define our full basis including the fermionic ones.
We do this in the following way: first we include the operators of Table 2 in ref.~\cite{EEMP2}.\footnote{Equivalently, for our discussion of bosonic operators, we could add the operators with fermions of Tables~2 and 3 in ref.~\cite{Grzadkowski:2010es}.} Now the set of operators is an over-complete basis since it contains $\text{\emph{20 bosonic operators}}+\text{\emph{44 operators with fermions}}=\text{\emph{64 operators}}$ in total. As shown in ref.~\cite{Grzadkowski:2010es}, the dim-6 basis contains a total of  $59$ operators (for a single family), therefore there are 5 redundant operators which we can remove. 
Performing field redefinitions, or equivalently using the EoM's, we can trade the three four-fermions operators of the first family
\bea
\label{red4F}
 (\bar{u}^1_R \gamma^\mu T^A u^1_R) (\bar{d}^1_R \gamma_\mu T^A d^1_R)\  , \quad (\bar{L}^1_L\sigma^a \gamma^\mu L^1_L)(\bar{L}^1_L \sigma^a \gamma_\mu L^1_L) \ , \quad   (\bar{e}^1_R \gamma^\mu e^1_R) (\bar{e}^1_R \gamma_\mu e^1_R) \ ,
 \label{RedcL3}
\eea
for \{${\cal O}_{2G}$, ${\cal O}_{2W}$, ${\cal O}_{2B}$\} of our basis and the operators of the first family
\bea
(i H^\dagger \sigma^a \lra{D}_\mu H)(\bar{L}^1_L \sigma^a \gamma^\mu L^1_L) \ , \quad  (i H^\dagger \lra{D}_\mu H)(\bar{e}^1_R  \gamma^\mu e^1_R) \ ,
 \label{RedcL3_2}
\eea
are removed in favour of the bosonic operators in Table~\ref{table:one1}, see Appendix~\ref{appen} for more details. This completes the definition of our dim-6 operator basis, for one family. In the present work,  we denote by $F$ and $f$  the fermion $SU(2)_L$ doublets and singlets, respectively, the subscripts $R$ or $L$ denote right or left-handed spinors. We put the  index \emph{i} (indistinguishably as a superscript or subscript) to denote either of the three SM families or, in some cases, to denote a particular SM fermion. Then, when convenient, we shall replace $F$ by $L$ or $Q$ to denote an $SU(2)_L$ lepton or quark doublet, respectively, and replace $f$ by either $\{e,u,d\}$ for the right-handed fermions of the first family, and so on for the other families. To generalize the basis to three families one has to add extra four-fermion operators and take into account the different flavor indices structures. Nonetheless, these extra operators do not affect our results. 

Let us comment on bases of common use in the literature. The set of operators 
\be
\{{\cal O}_W, {\cal O}_B,{\cal O}_{WW},{\cal O}_{WB},{\cal O}_{BB}\}
\ee
 is in one-to-one correspondence with the operators used in ref.~\cite{Hagiwara:1993ck}
 \be
 \{{\cal O}_{HW}, {\cal O}_{HB},{\cal O}_{WW},{\cal O}_{WB},{\cal O}_{BB}\} \ ,
 \ee where $ {\cal O}_{HW}\equiv ig (D^\mu H)^\dagger \sigma^a (D^\nu H) W^a_{\mu\nu}  ,  \ {\cal O}_{HB}\equiv ig^\prime(D^\mu H)^\dagger  (D^\nu H) B_{\mu\nu} $, and with the ones used in ref.~\cite{Giudice:2007fh}
  \be
 \{{\cal O}_{W}, {\cal O}_{B},{\cal O}_{HW},{\cal O}_{HB},{\cal O}_{BB}\} \ .
 \ee 
Our basis has the advantage that the anomalous dimension matrix of the sector $\{{\cal O}_{B}, {\cal O}_{W}\}\times\{{\cal O}_{BB},{\cal O}_{WB}, {\cal O}_{WW}\}$ is block diagonal~\cite{Elias-Miro:2013gya}. The anomalous dimension in the other bases is given in Appendix~\ref{appenHagiwara}. As the SILH basis~\cite{Giudice:2007fh}, our basis also separates the operators generated at tree-level from the ones obtained at the radiative level only, when the new physics degrees of freedom, assumed to be weakly coupled, are integrated out~\cite{Elias-Miro:2013gya}. When the Higgs emerges as pseudo Nambu--Goldstone boson, the SILH basis further makes the distinction between a loop involving new-physics interactions and a loop involving SM interactions only.

In this paper, we are limiting ourselves to the set, ${\cal B}_1$, of 13 operators appearing in Table~\ref{table:one1} (omitting ${\cal O}_6$ that does not contribute directly to the 13 physical observables we are studying). We compute the running of ${\cal B}_1$ into ${\cal B}_1$. If the remaining set of independent operators, needed to complete the basis specified above, is denoted by ${\cal B}_2$, there could also be  {\it i)} a running of ${\cal B}_2$ into ${\cal B}_1$, {\it ii)} a running of ${\cal B}_1$ into ${\cal B}_2$ and of course {\it iii)} a running of ${\cal B}_2$ into itself. The first effect would reflect itself in new RG contributions to our list of low-energy observables; under our hypothesis of no-tuning (or no correlations) among the different RG contributions these effects do not change our RG-induced bounds on the operators in $\mathcal{B}_1$. In principle new RG-induced bounds on some operators in $\mathcal{B}_2$ could be obtained, however we already commented on the fact that this is not the case for $\Op_6$ and $\Op_{y_{u,d,e}}$.
The second effect could, in principle, allow us to obtain new RG-induced bound on the operators in $\mathcal{B}_1$ via the mixing to some tightly constrained operators in $\mathcal{B}_2$, for example via the mixing to $\Op_L$ and $\Op_{LL}^{12}$, as we mentioned above. 
The study of these effects would be an interesting generalization of our ideas but would require the computation of the full anomalous dimension matrix and a complete phenomenological analysis of all the observables relevant to the dimension-6 operators, which is beyond the purpose of the present work.

\section{One-loop scaling of EW and Higgs operators}
\label{oneLoopRGES}

In general, quantum effects mix all the operators among themselves when going from the scale of new physics down to the scale at which the experimental measurements are performed. However, the 3 operators with gluons, ${\cal O}_{GG}, {\cal O}_{2G}$ and ${\cal O}_{3G}$, constitute a separate sector that does not mix with the other 11 bosonic operators at one-loop.\footnote{The only exception is a contribution from $\Op_{2B}$ to the RG of $\Op_{2G}$, see Table~\ref{tab:OurBasisMatrixGluons}. This mixing, however, is phenomenologically not very relevant since the Wilson coefficient of $\Op_{2B}$ is strongly constrained, as we show in Section~\ref{constraints}. In Section~\ref{GluonSect} we present the anomalous dimension of the three operators with gluons.} So, even if ${\cal O}_{GG}$ affects Higgs physics by controlling the dominant production mode of the Higgs boson at the LHC, it can be treated separately from the 3 other Higgs observables we are interested in here.
Furthermore since the Higgs self-interactions have not been measured yet, and since ${\cal O}_6$ does not enter into the anomalous dimensions of any dim-6 operator other than itself, it can also be omitted from our analysis. For the Higgs- and EW-sector RG study, we can thus restrict to the following set of 10 dim-6  operators and compute the corresponding anomalous dimension matrix
\begin{equation}
\left\{{\cal{O}}_H, {\cal{O}}_T, {\cal{O}}_B, {\cal{O}}_W, {\cal{O}}_{2B}, {\cal{O}}_{2W}, {\cal{O}}_{BB}, {\cal{O}}_{WW}, {\cal{O}}_{WB}, {\cal{O}}_{3W}\right\} \ .
\label{univBasis}
\end{equation}
We include all the one-loop contributions proportional to $c_i$ and depending on 
\begin{equation}
\left\{ g^{\prime},\, g,\, g_s,\, \lambda,\, y_t \right\} \, ,
\label{orderCalc}
\end{equation}
where $g^{\prime}$, $g$ and $g_s$ are the respective $U(1)_Y$, $SU(2)_L$ and $SU(3)_c$ gauge couplings, $\lambda$ is the Higgs quartic coupling and $y_t$ is the Yukawa coupling of the top quark, i.e. we neglect the contributions proportional to the Yukawas of the light fermions ($y_b/y_t \sim 0.02$, $y_b$ is the bottom quark Yukawa). The couplings are normalised such that
\begin{equation}
{\cal L}_{\text{SM}}= {\cal L}_{\text{Kin}}  +m^2|H|^2 -\lambda \left| H\right|^4  - y_t \left[\bar{Q}_L t_R \tilde{H} +\text{h.c.}  \right] + {\cal O}\left( y_l\right )\, ,
\end{equation}
where $\tilde H = i \sigma_2 H^*$, $Q_L$ is the third family quark doublet $\left(t_L, b_L\right)^T$ whose weak hypercharge is $Y_{L}=1/6$, $t_R$ is the right-handed top quark of weak hypercharge $Y_R=2/3$ and $y_l$ denotes the Yukawa couplings of the fermions lighter than the top. The kinetic term ${\cal L}_{\text{Kin}}$ contains covariant derivatives, defined in the previous section, that determine the couplings of the Higgs doublet and fermions to the gauge bosons as well as gauge bosons self-interactions. 

We regularized the loop integrals using dimensional regularisation and used $\overline{\text{MS}}$ subtraction scheme. We  performed the computation in the unbroken phase of the SM and in the background field gauge, with the gauge fixing term
\begin{equation}
{\cal L}^{g.f.} = - \frac{1}{2 \xi_A} (D^{(A)}_\mu \delta A^{a \mu})^2 \, ,
\label{bkgFG}
\end{equation}
 where $\delta A=\left\{\delta B, \delta W, \delta G\right\}$ is the quantum field with respect to which the $\text{dim}\geq4$ SM action is path-integrated and $D^{(A)}_\mu$ is the covariant derivative with respect to the corresponding background field $A=\left\{ B,  W,  G\right\}$.
 
 In Table~\ref{tab:OurBasisMatrix}, we give the one-loop anomalous dimensions of the operators of ~\eq{univBasis}, in the basis defined in Section~\ref{Dim6Basis}.\footnote{The self-renormalization of $c_{3W}$ has been extracted from the computation of refs.~\cite{Narison:1983,Morozov:1985ef}, where the authors calculated the one of $c_{3G}$.}  We have defined 
\be
\gamma_{c_i}= 16\pi^2 \frac{d c_i}{d \log \mu}\ .
\ee 
A common effect encountered while computing the  RG scaling of the above operators is the appearance of counter-terms which correspond to dim-6 operators that are  not in our basis (the computation \emph{does not know} our choice of basis) \cite{EEMP2}.  These radiatively-generated redundant operators need to be redefined into operators present in our basis. Upon redefinition, these redundant  operators contribute to the anomalous dimensions of the operators in our basis at the same order as other direct contributions coming from one-particle-irreducible graphs. For details on the radiatively generated operators and how we deal with the redundant ones, see Appendix~\ref{appen}.  Notice that the matrices of Table~\ref{tab:OurBasisMatrix} already contain these indirect effects. This ensures that the result is gauge invariant and indeed we checked that the result is independent of the gauge fixing parameters $\xi_A$ of \eq{bkgFG}. 


{
\renewcommand{\arraystretch}{1.8} 
\renewcommand{\tabcolsep}{2mm}
\begin{table}[thpd]
\begin{center}
\scriptsize
\begin{tabular}{c|cc}
 &   $c_{H}$ &  $c_{T}$ \\
\hline 
 $\gamma_{c_{H}}$ &  $-\frac{9}{2}g^{2}-3g^{\prime2}+24\lambda+12y_{t}^{2}$ &  $-9g^{2}+\frac{9}{2}g^{\prime2}+12\lambda$ \\
 $\gamma_{c_{T}}$ &  $\frac{3}{2}g^{\prime2}$ &  $\frac{9}{2}g^{2}+12\lambda+12y_{t}^{2}$ \\
 $\gamma_{c_{B}}$ &  $-\frac{1}{3}$ &  $-\frac{5}{3}$ \\
$\gamma_{c_{W}}$  &  $-\frac{1}{3}$ &  $-\frac{1}{3}$ \\
other $\gamma_{c_{i}}$'s &  $0$ or $ {\cal O}(y_l)$ &  $0$  or ${\cal O}(y_l)$ 
\end{tabular}

\vspace{0.5cm}

\begin{tabular}{C{0.8cm}|cccc}
 &   $c_{B}$ &  $c_{W}$ &  $c_{2B}$ &  $c_{2W}$ \\ \hline
 $\gamma_{c_{H}}$ &  $-\frac{9}{4}g^{\prime2}(g^{\prime2}-2g^{2})-6\lambda g^{\prime2}$ &  $\frac{9}{4}g^{2}(2g^{\prime2}-g^{2})-36\lambda g^{\prime2}$ & $-\frac{141}{16}g^{\prime4}+3g^{\prime2}\lambda$  &  $\frac{63}{8}g^{4}+\frac{51}{16}g^{2}g^{\prime2}+18\lambda g^{2}$  \\
 $\gamma_{c_{T}}$  &  $-\frac{9}{4}g^{\prime2}g^{2}-6\lambda g^{\prime2}$ &  $-\frac{9}{4}g^{\prime2}g^{2}$ &  $3g^{\prime4}+\frac{9}{8}g^{\prime2}g^{2}+3\lambda g^{\prime2}$  &  $\frac{9}{8}g^{\prime2}g^{2}$  \\
 $\gamma_{c_{B}}$  &  $\frac{g^{\prime2}}{6}+6y_{t}^{2}$ &  $\frac{g^{2}}{2}$ & $\frac{59}{4}g^{\prime2}$ &  $-\frac{g^{2}}{4}$  \\
 $\gamma_{c_{W}}$  &  $\frac{g^{\prime2}}{6}$ &  $\frac{17}{2}g^{2}+6y_{t}^{2}$ &  $\left(\frac{29}{8}-\frac{53 g^{\prime2}}{4g^{2}}\right)g^{\prime2}$  &  $\frac{79}{8}g^{2}+\frac{29}{4}g^{\prime2}$  \\
 $\gamma_{c_{2B}}$  &  $-\frac{2}{3}g^{\prime2}$ &  $0$  &  $\frac{94}{3} g^{\prime2}$  &  $0$  \\
 $\gamma_{c_{2W}}$  &  $0$ &  $-\frac{2}{3}g^{2}$  &  $\left(\frac{53}{12}-\frac{53 g^{\prime2}}{4g^{2}}\right)g^{\prime2}$ &  $\frac{331}{12}g^{2}+\frac{5}{4}g^{\prime2}$  \\
 $\gamma_{c_{BB}}$  &  $0$ &  $0$ &  $0$  &  $0$ \\
 $\gamma_{c_{\text{\emph{WW}}}}$  &  $0$ &  $0$  &  $0$  &  $0$ \\
 $\gamma_{c_{WB}}$  &  $0$ &  $0$ &  $0$  &  $0$  \\
 $\gamma_{c_{3W}}$  &  $0$  &  $0$  &  $0$  &  $0$ 
\end{tabular}

\vspace{0.5cm}

\begin{tabular}{C{0.8cm} | C{3.115cm} C{3.115cm} C{3.115cm} C{3.115cm} }
                                        &  $c_{BB}$ & $c_{WW}$& $c_{WB}$&$c_{3W}$  \\ \hline
 $\gamma_{c_H}$     &  $0$&$0$&$0$  &  $0$      \\ 
 $\gamma_{c_T}$            &$0$&$0$&$0$ &  $0$  \\ 
$\gamma_{c_B}$           & $0$ & $0$ & $0$ & $0$      \\ 
 $\gamma_{c_W}$       &      $0$ & $0$ & $0$ &  $0$     \\ 
 $\gamma_{c_{2B}}$      &$0$&$0$& $0$ & $0$   \\ 
 $\gamma_{c_{2W}}$    &$0$&$0$&$0$ & $0$    \\ 
 $\gamma_{c_{BB}}$     &$\frac{g^{\prime 2}}{2}-\frac{9g^2}{2}+6y_t^2+12\lambda$ & $0$ & $3 g^2$ &$0$\\ 
 $\gamma_{c_{\text{\emph{WW}}}}$  &  $0$ & $-\frac{3 g^{\prime 2}}{2} - \frac{5 g^2}{2}+6y_t^2+12\lambda$ & $g^{\prime 2} $ & $ \frac{5}{2} g^2$ \\ 
 $\gamma_{c_{WB}}$   & $2 g^{\prime 2}$  & $2 g^2$ & -$\frac{g^{\prime 2}}{2}+\frac{9g^2}{2}+6y_t^2+4\lambda$ &$-\frac{g^2}{2}$ \\ 
   $\gamma_{c_{3W}}$   &$0$&$0$&$0$    & $\frac{53}{3} g^2$
 \end{tabular}

\caption{\small Anomalous dimension matrix for the Wilson coefficients of the dim-6 bosonic operators, in the basis defined in Section~\ref{Dim6Basis}.}
\label{tab:OurBasisMatrix}
\end{center}
\end{table}
}

 
Apart from gauge invariance, there is another non-trivial consistency check that we have performed. The current-current operators in the left box  of Table~\ref{table:one1} can be related to each other and to other current-current operators containing fermions by using the SM EoM, or equivalently by carrying out field redefinitions. In a hypothetical theory without fermions\footnote{The anomalous dimension matrix of this fermionless theory is related, though not equal, to the anomalous dimension matrix we have computed, that is why considering this hypothetical theory provides a non-trivial test of our computation.}, some contributions of the operators in the left box of Table~\ref{table:one1} would vanish upon the EoM, i.e. they would form an over-complete set of operators. This would also imply relationships between independently computed   entries in the anomalous dimension matrix or, in other words, the anomalous dimensions of this over-complete set is invariant under changes in the field coordinates that respect the SM gauge symmetries. Our matrix passes this consistency check as we shall discuss in detail in Appendix~\ref{appenB}. We emphasize that the set of 59 operators introduced in Section \ref{Dim6Basis} is a basis, i.e. it does not contain any redundant operators; it is over-complete only in the hypothetical theory without fermions.

Some parts of the anomalous dimension matrix presented here, have been calculated in previous literature~\cite{Narison:1983,Morozov:1985ef,Hagiwara:1993ck,Hagiwara:1994pw,Alam:1997nk,Barbieri:2007bh,Grojean:2013kd,Elias-Miro:2013gya,EEMP2,Jenkins:2013zja,Jenkins:2013wua}. In some cases these previous computations use methods different from ours, but we find complete agreement in the final results. We present a detailed comparison with previous literature, including a discussion about the difference in our methods in Appendix~\ref{appenHagiwara}.

\section{RG-induced contraints on EW and Higgs observables}
\label{Sconst}

In this section we discuss the possibility to use the RGE's to derive constraints on the Wilson coefficients at the weak scale by requiring that none of the RG contributions to these weak-scale Wilson coefficients exceeds the direct bounds \cite{Hagiwara:1993ck}. 
Since the RGE's mix various operators, it becomes possible to put tight constraints on operators loosely constrained by direct measurements via their RG contributions to more severely constrained operators. Then, in Section~\ref{constraints}, we apply our method and use EW precision data, triple gauge couplings measurements and Higgs data to derive RG-induced bounds on the set of 10  observables we are interested in. 

Renormalizing, order by order, the effective action, the logarithmically divergent terms computed in the previous section are absorbed in the definition of the renormalized Wilson coefficient. If one is interested in obtaining bounds on the Wilson coefficients at the low scale $\sim m_H$, the only effect of the 1-loop diagrams are small finite terms, proportional to $\sim \log m_H / m_Z$, which we did not compute here. Allowing for arbitrary cancellations in the definition of the renormalized coefficients renders the 1-loop effects small and the indirect bounds which can be obtained in this way are quite weak \cite{Mebane:2013zga} and not competitive with direct bounds from Higgs physics and anomalous TGC measurements.
We follow a different approach, already outlined in ref.~\cite{Hagiwara:1993ck}. We are interested in obtaining indirect bounds on the UV value of the Wilson coefficients from low-energy experiments, in this case the 1-loop effect is enhanced by $\sim \log \Lambda / m_H$. Moreover, we assume that no tuned cancellations (or correlations) are present in the definition of the renormalized coefficients and require each log-divergent term not to exceed the direct bounds. In this way, our indirect bounds are much stronger than in ref.~\cite{Mebane:2013zga} and, more importantly, are useful in order to obtain insight into the UV physics. In fact, if any of our RG-induced bounds would be violated by a direct measurement this would imply a particular pattern of cancellation (or correlation) in the UV dynamics.

\subsection{How much fine-tuning is needed to accommodate the data?}  
\label{tuning}

The electroweak and Higgs observables we are interested in (specified in Section~\ref{constraints}) receive contributions from a particular linear combination of the dim-6 operator's Wilson coefficients, suitably multiplied by the SM couplings:
\be
	(\text{obs})_i=\kappa_i+\omega_{ij} c_j \equiv \kappa_i + \hat{c}_i\ \quad \rightarrow \quad
	\delta (\text{obs})_i = \hat{c}_i \ ,
	\label{omegaObs}
\ee
where $\kappa_i$ is the SM contribution, the $c_k$'s are the Wilson coefficients and $\omega_{ij}$ is a matrix containing the SM couplings and ratios of scales ($\omega \sim {\cal O}(m_W^2 / \Lambda^2)$). We defined $\hat{c}_i$ as the linear combinations of the Wilson coefficients which contribute directly to each observable $(\text{obs})_i$ and we shall refer to them in the following as \emph{observable couplings}, with a slight abuse of language. If the new combinations $\hat{c}_i$ are independent, this corresponds to a change of basis such that to each operator corresponds an observable; we shall call this the \emph{observable basis}.

As an example, consider the process $h \rightarrow \gamma Z$ which receives a contribution from the SM (in this case at one loop) as well as a direct contribution from a linear combination of the dim-6 operators. We parametrize this contribution with the \emph{observable coupling} $\hat{c}_{\gamma Z}$, to be defined in \eq{higgsL}, which is related to the Wilson coefficients of our basis as ($c_{\theta_W}$ and $s_{\theta_W}$ are respectively the sinus and cosinus of the weak mixing angle ${\theta_W}$)
\be
	\hat{c}_{\gamma Z} =\, \frac{m_W^2}{\Lambda^2} \left( 2 c_{\theta_W}^2c_{WW}- 2 s_{\theta_W}^2 c_{BB}-(c_{\theta_W}^2-s_{\theta_W}^2)c_{WB} \right) \ .
	\label{eq:ObservableExample1}
\ee
The above relation defines the coefficients $\omega_{\gamma Z, j}$ for this particular observable.

Now, suppose that this set of observables receives lower and upper bounds from experimental measurements: 
\be
	\delta (\text{obs})_i|_{m_h} = \hat{c}_i (m_h) = \omega_{ij}(m_h) c_j(m_h) \in [ \epsilon_i^{low}, \epsilon_i^{up} ]   \ .
	\label{eq:directConstraint}
\ee
The observable coupling $\hat{c}_i(m_h)$ (constrained at low energy) is related, through the running, to the high-scale value of the Wilson coefficients $c_j(\Lambda)$, which is not directly known since it is determined by the BSM degrees of freedom that have been integrated out.
The matrix $\omega_{ij}(m_h) $ also runs with the scale (in the example of \eq{eq:ObservableExample1} this would be the running of $g, g^\prime$ and $v$ inside $m_W$ and $\theta_W$), however we are not interested in such a running because $\omega_{ij}$ is determined by measurements performed at the EW scale and because, for the purpose of this work, we are not interested in the UV value of the SM couplings. This is the reason why we have not taken care of the contributions of the dim-6 operators on the SM couplings, parametrized by $\kappa_i$ in \eq{omegaObs}, which would only be necessary if we wanted to relate $\omega_{ij}(m_h)$ to $\omega_{ij}(\Lambda)$ at the order we are working.

This discussion leads us to define the scale-dependent observable couplings as 
\be
\hat{c}_i(\mu) \equiv \omega_{ij}(m_h) c_j(\mu) \ ,
\ee 
obtaining
\be
\delta (\text{obs})_i|_{m_h} = \hat{c}_i(m_h) =  \hat{c}_i(\Lambda)-\frac{1}{16\pi^2} \hat{\gamma}_{ij}\hat{c}_j(\Lambda) \log\left(\frac{\Lambda}{m_h}\right) \ ,
\label{ObsCoeffRunning}
\ee
where
\be
\hat{\gamma}_{ij} \equiv \omega_{ik}(m_h) \, \gamma_{kl} \, \omega^{-1}_{lj}(m_h)
\ee
and $\gamma_{kl}$ is the matrix computed in the previous section. Our interest in \eq{ObsCoeffRunning} is twofold: we want to find instances where a less constrained operator can mix with a more constrained one by appearing in its RGE's and secondly (but closely related),  to learn about the new degrees of freedom at the matching scale. In the following we shall work at leading-log order, which is fine if the hierarchy between the new physics scale $\Lambda$ and the EW scale is not too big.

The fundamental assumption we make in order to obtain an indirect constrain on the $\hat{c}_j(m_h)$ through the RG is that we require each term in the sum on the r.h.s. of \eq{ObsCoeffRunning}, proportional to some coefficient $\hat{c}_j$, to be contained in the experimental bounds associated to the observable $\delta (\text{obs})_i|_{m_h}$:
\begin{gather}
	(1 - \delta_i) \hat{c}_i (\Lambda) \in [ \epsilon_i^{low}, \epsilon_i^{up} ]  \, ,
	\label{eq:MatchingBound1} \\
	-\frac{1}{16\pi^2} \hat{\gamma}_{i\hat{\jmath}}\hat{c}_{\hat{\jmath}}(m_h) \log\left(\frac{\Lambda}{m_h}\right) \in [ \epsilon_i^{low}, \epsilon_i^{up} ]  \, ,
	\label{eq:IndirectBound1}
\end{gather}
where we defined $\delta_i = \hat{\gamma}_{ii} / (16\pi^2) \log(\Lambda / m_h)$ and in the last line the index $\hat{\jmath}$ is not summed over.\footnote{In the following we shall denote with a hat all repeated indices which are not summed over.} We have also used the fact that substituting $\hat{c}_j(\Lambda)$ for $\hat{c}_j(m_h)$ in the $\hat{\gamma}_{ij}\hat{c}_j$ term of \eq{ObsCoeffRunning} amounts to corrections ${\cal O}\left((4\pi)^{-4} \log^2(\Lambda/m_h)\right)$ that are beyond our precision (the same is true for the evaluation of $\gamma_{ij}$). Notice that this assumption is not only a requirement of the absence of fine-tuning but also an hypothesis on the UV physics, since particular relations, due to symmetry or dynamical accidents, between those combinations could be generically found when considering a BSM theory. From our bottom-up approach we parametrize also this absence of correlations as an absence of tuning.
From \eq{eq:MatchingBound1} we can put bounds on the matching-scale Wilson coefficients $c_j(\Lambda)$:
\be
	c_j(\Lambda) \in \, \left[ \sum_i (1 - \delta_i)^{-1} \omega_{ji}^{-1} \epsilon_i^{low}, \sum_i  (1 - \delta_i)^{-1} \omega_{ji}^{-1}\epsilon_i^{up} \right] \ ,
	\label{eq:MatchingBound2}
\ee
notice that, as expected, they grow quadratically weaker with the increase of the UV scale $\Lambda$ since $\omega^{-1} \sim \Lambda^2 / m_W^2$. 
Using \eq{eq:IndirectBound1}, instead, we can put an RG-induced bound on the observable $\delta (\text{obs})_j|_{m_h}$ using the direct constraints on $\delta (\text{obs})_i|_{m_h}$, \eq{eq:directConstraint}:
\be\begin{split}
	\text{if }\hat{\gamma}_{\hat{\imath} j}>0: \quad&  \delta (\text{obs})_j |_{m_h} \in \frac{16\pi^2}{ \log \left(\Lambda / m_h \right)} (\hat{\gamma}_{\hat{\imath} j})^{-1} [-\epsilon_{\hat{\imath}}^{up}, - \epsilon_{\hat{\imath}}^{low}] \ , \\
	\text{if }\hat{\gamma}_{\hat{\imath} j}<0: \quad& \delta (\text{obs})_j |_{m_h} \in \frac{16\pi^2}{ \log \left(\Lambda / m_h \right)} (\hat{\gamma}_{\hat{\imath} j})^{-1} [ \epsilon_{\hat{\imath}}^{low}, \epsilon_{\hat{\imath}}^{up}] \, .
	\label{eq:RGinducedBounds}
\end{split}\ee
The indirect bounds in \eq{eq:RGinducedBounds}, grow logarithmically stronger with the increase of the UV scale $\Lambda$. However, since the expected effects from new physics decrease quadratically with $\Lambda$, assuming order one coefficients $c_i$, even if the RG-induced bounds on the observables become slightly stronger, their power in investigating the UV degrees of freedom becomes much weaker for higher values of $\Lambda$, as is clear from \eq{eq:MatchingBound2}.
It might seem that these bounds are not significant because of the  loop factor in the above equation; all the $\epsilon_i$'s are, however, not of the same order and if  $| \epsilon_i^{low,up} | \ll | \epsilon_j^{low,up} |$, the bound in the above equation can be stronger than the direct bound on  $\delta (\text{obs})_j |_{m_h}$, in spite of the  loop factor. The RG-induced bounds are, thus, significant only when a weakly constrained coupling appears in the RGE of a strongly coupled one.

Once new physics effects will be, hopefully, observed and the constraints of \eq{eq:directConstraint} will not include the zero value in the allowed interval ($0 < \epsilon^{low}_i< |\delta (\text{obs})_i|_{m_h} < \epsilon^{up}_i$), another interesting information that could be extracted from RG effects is a quantification of how much tuned, among themselves, are the electroweak and Higgs observables. First of all, let us define the fine-tuning in an observable as~\cite{Barbieri:1987fn}
\be
 \Delta_i \equiv \text{Max}_j \left| \frac{\partial \log \delta (\text{obs})_i|_{m_h}}{ \partial \log \hat{c}_j(\Lambda)} \right| 
 	\simeq \text{Max} \left\{ \frac{ |\hat{c}_i(\Lambda) |}{| \delta (\text{obs})_i|_{m_h}}, \, \frac{ \log\left(\Lambda / m_h \right)}{16\pi^2} \frac{ \text{Max}_{j\neq i} \left|  \hat{\gamma}_{i\hat{\jmath}}\right| |\delta (\text{obs})_{\hat{\jmath}}|_{m_h}}{| \delta (\text{obs})_i|_{m_h}} \right\}   \ ,
 \label{eq:tuning}
\ee
where in the second step we separated the diagonal contribution from the off-diagonal ones and, for the diagonal term, we neglected the loop contribution since $\hat{c}_i(\Lambda)$ enters already at tree level and this would be its leading contribution to the tuning. In particular, the fine-tuning $\Delta_i$ will satisfy,
\be
	 \Delta_i \geq \frac{ \log\left(\Lambda / m_h \right)}{16\pi^2} \frac{ \text{Max}_{j\neq i} \left|  \hat{\gamma}_{i\hat{\jmath}}\right| |\delta (\text{obs})_{\hat{\jmath}}|_{m_h}}{| \delta (\text{obs})_i|_{m_h}}
	 > \frac{ \log\left(\Lambda / m_h \right)}{16\pi^2} \frac{ \text{Max}_{j\neq i} \left|  \hat{\gamma}_{i\hat{\jmath}}\right| \epsilon_{\hat{\jmath}}^{low}}{\epsilon_i^{up}},
	  \label{eq:tuning2}
\ee
and one might be able to conclude that a certain degree of fine-tuning among the contributions to the RG flow of some operator is necessary.

\subsection{EW and Higgs observables}
\label{constraints}

Let us now apply the general formulas of the previous section to the electroweak and Higgs observables we want to constrain. In Section~\ref{Dim6Basis} we have considered 10 EW and Higgs operators
\be
{\cal O}_H, \ {\cal O}_T, \ {\cal O}_{W}, \ {\cal O}_B, \ {\cal O}_{2W}, \ {\cal O}_{2B}, \ {\cal O}_{WW}, \ {\cal O}_{WB}, \ {\cal O}_{BB}, \ {\cal O}_{3W}, \  
\label{set10}
\ee
 to parametrize BSM corrections to the SM Lagrangian. Let us now describe in detail the set of pseudo-observables, briefly mentioned in Section~\ref{Dim6Basis}, that constrain all these operators and form our observable basis.  These include the four electroweak oblique parameters $\hat{S}$, $\hat{T}$, $Y$ and $W$, constrained by LEP 1 and LEP 2, the three anomalous triple gauge coupling (TGC) and three observables related to Higgs physics: the decays to $\gamma\gamma$, $\gamma Z$ and a universal rescaling of all the branching ratios \cite{Pomarol:2013zra}.  To derive the RG-induced constraints on these observables we first need to relate them to the operators in \eq{set10}, that is define the transformation matrix, $\omega_{ij}$,  from the basis in \eq{set10} and to the observable basis.
 
We begin with the electroweak precision observables constrained by measurements at LEP1, LEP2 and Tevatron.  The first step of the analysis is  to fix the SM parameters $g$, $g^\prime$  and $v$  by the three most precise measurements: the Fermi constant $G_F$ in muon decays, the fine-structure constant $\alpha_{em}$ and the $Z$-boson mass $m_Z$. With the input parameters fixed, the  SM gives predictions for observables such as  $Z$-pole measurements at LEP 1, the Tevatron measurement of the  $W$-mass and LEP 2 measurements of the $e^+e^- \to f^+f^-$ cross-sections. New physics can affect this analysis by either changing the relationship between    the input parameters $g$, $g^\prime$  and $v$ to the measurement of $G_F$,  $\alpha_{em}$ and  $m_Z$ or by directly contributing to the other measurements. All the deviation in the above observables induced by the operators we consider, \eq{set10}, can be parametrized by the $\hat S$, $\hat T$, $W$ and $Y$ parameters~\cite{Barbieri}
\be
\Delta \lag_{\text{EWPT}} =  -\frac{\hat{T}}{2}\frac{m_Z^2}{2} Z_\mu Z^\mu - \frac{\hat{S}}{4 m_W^2} \frac{ g g' v^2}{2}(W^3_{\mu \nu}B^{\mu \nu})-\frac{W}{2 m_W^2}   ( \partial^\mu  W_{\mu \nu}^3)^2-\frac{Y}{2 m_W^2}  (  \partial^\mu  B_{\mu \nu})^2.
\label{ewpt}
\ee
 The contribution  of the Wilson coefficients of the operator set in \eq{set10} to the above observables is given by, 
 \bea 
&& \hat{T}=\hat{c}_T (m_W) = \frac{v^2}{\Lambda^2} c_T(m_W) \ , \quad 
	\hat{S}=\hat{c}_S (m_W)=  \frac{m_W^2}{\Lambda^2} \left[c_W(m_W) + c_B(m_W) + 4 c_{WB}(m_W)\right], \nonumber\\
&& Y=\hat{c}_{Y}(m_W)  = \frac{m_W^2}{\Lambda^2} c_{2B}(m_W) \ , \qquad
	W=\hat{c}_{W}(m_W) = \frac{m_W^2}{\Lambda^2} c_{2W}(m_W) \  . 
 \label{eq:ObsCoeffEWPT} 
 \eea
The above parameters have been measured very precisely and are constrained at the per mille level. We present the 95 $\%$ CL bounds on these parameters   in Table~\ref{table:ObsBasisNum}. 

A second set of independent measurements that constrain the operator set in \eq{set10} are the TGC that were measured in the $e^+e^- \to W^+ W^-$ process at LEP2. The phenomenological Lagrangian to describe  deviations in the TGC observables, from their SM values, is \footnote{Note that in the previous version of the paper  the deformations related to $\delta g^1_Z$ and $\delta \kappa_\gamma$ were defined with a sign opposite to that used in the literature. We have changed this, and now we use the conventional definitions.} 
\bea
	\Delta \lag_{3V} &=&\; i g \, g_1^Z c_{\theta_W} Z^\mu \left( W^{+\nu} \hat{W}^-_{\mu\nu}  - W^{-\nu} \hat{W}^+_{\mu\nu} \right) + i g \left( \kappa_z c_{\theta_W} \hat{Z}^{\mu\nu} + \kappa_\gamma s_{\theta_W} \hat{A}^{\mu\nu} \right) W^+_\mu W^-_\nu \nonumber\\
		&+ &\frac{ig}{m_W^2} \left( \lambda_Z c_{\theta_W} \hat{Z}^{\mu\nu} + \lambda_\gamma s_{\theta_W} \hat{A}^{\mu\nu} \right) \hat{W}^{-\rho}_\mu \hat{W}^+_{\rho\nu},
		\label{eq:3GVLagr}
 \eea
where $\hat{V}_{\mu\nu} = \partial_\mu V_\nu - \partial_\nu V_\mu$, the photon field $A_\mu=c_{\theta_W} B_\mu+s_{\theta_W}  W^3_\mu$ has field-strength $\hat{A}_{\mu\nu}$, while 
$Z_\mu=c_{\theta_W}  W^3_\mu-s_{\theta_W} B_\mu$ has field-strength $\hat{Z}_{\mu\nu}$ and we use $s_{\theta_W}\equiv \sin\theta_W=g^\prime/\sqrt{g^2+g^{\prime 2}}$, $c_{\theta_W}\equiv \cos\theta_W=g/\sqrt{g^2+g^{\prime 2}}$ and $e=gs_{\theta_W}$. 
Note that the above Lagrangian has only three independent parameters at the dim-6 level  taken to be  $ g_1^Z, \kappa_\gamma$  and $\lambda_\gamma$ here; the other two can be expressed as : $\lambda_Z = \lambda_\gamma$ and $\kappa_Z =  g_1^Z - t^2_{\theta_W} \kappa_\gamma$. These relations are a consequence of the accidental custodial symmetry that is preserved by the dim-6 operators entering in the TGC~\cite{Gounaris:1996rz}.
The SM contribution is given by $(g_1^Z)_{SM} =  (\kappa_\gamma)_{SM} = 1$ and $ (\lambda_Z)_{SM} = 0$. The corrections induced by the dim-6 operators in our basis are given by:
\be \begin{split}
	& \delta g_1^Z \equiv \hat{c}_{gZ}  (m_W)= - \frac{m_W^2}{\Lambda^2} \frac{1}{c_{\theta_W}^2} c_W (m_W) \ , \qquad
	\delta \kappa_\gamma \equiv \hat{c}_{\kappa\gamma} (m_W) = \frac{m_W^2}{\Lambda^2} 4 c_{WB}(m_W) \ , \\
	& \lambda_Z \equiv \hat{c}_{\lambda \gamma} (m_W) = - \frac{m_W^2}{\Lambda^2} c_{3W}(m_W) \ ,
\end{split}  \label{eq:ObsCoeffTGC}  \ee
where $\delta g_1^Z = g_1^Z - (g_1^Z)_{SM}$ and $\delta \kappa_\gamma = \kappa_\gamma - (\kappa_\gamma)_{SM}$.
 The constraints on these TGC observables are at the percent level (see Table~\ref{table:ObsBasisNum}) and thus at least an order of magnitude weaker than the constraints on the electroweak parameters in \eq{eq:ObsCoeffEWPT}. Note  that, for this reason,  in \eq{eq:ObsCoeffTGC} we have ignored contributions to the  $e^+e^- \to W^+ W^-$ process from the couplings in \eq{ewpt}.

Higgs physics provides the three  remaining observables for our observable basis. We consider the  branching ratios $h\rightarrow \gamma \gamma / Z \gamma$ and the correction to the Higgs kinetic term,
\be
	\Delta \lag_{Higgs} \supset \frac{\hat{c}_H}{2}\frac{(\partial_\mu h)^2}{2}+ \frac{\hat{c}_{\gamma\gamma} e^2}{m_W^2}\frac{h^2}{2}\hat{A}_{\mu \nu}\hat{A}^{\mu \nu}+ \frac{\hat{c}_{\gamma Z}~e g }{m_W^2 c_{\theta_W}} \frac{h^2}{2} \hat{A}_{\mu \nu}\hat{Z}^{\mu \nu}.
	\label{higgsL}
\ee
The above coefficients, in terms of the dim-6 operator's Wilson coefficients are given by
\be \begin{split}
	\hat{c}_H (m_h) &=\, \frac{v^2}{\Lambda^2} c_H (m_h), \qquad \\
	\hat{c}_{\gamma\gamma}(m_h) &=\, \frac{m_W^2}{\Lambda^2} \left( c_{BB}(m_h)+c_{WW}(m_h)-c_{WB}(m_h) \right), \qquad \\
	\hat{c}_{\gamma Z}(m_h) &=\, \frac{m_W^2}{\Lambda^2} \left( 2 c_{\theta_W}^2c_{WW}(m_h)- 2 s_{\theta_W}^2 c_{BB}(m_h)-(c_{\theta_W}^2-s_{\theta_W}^2)c_{WB}(m_h) \right).
\end{split}  \label{eq:ObsCoeffHiggs} \ee
We present the constraints on these three observables in Table~\ref{table:ObsBasisNum}. The coupling $ \hat{c}_{\gamma \gamma}$ is constrained at the per mille level although the constraint on the SM diphoton width has been measured only with  ${\cal O}(1)$ precision. This is because the SM width is already one-loop suppressed and thus the current ${\cal O}(1)$ precision of measurement corresponds  to $\hat{c}_{\gamma\gamma}\approx 10^{-3}$.  The correction to the Higgs kinetic term $\hat{c}_H$ on the other hand is  poorly constrained. This is because $\hat{c}_H$ causes a universal shift in all the Higgs couplings and thus drops out from the branching ratios. Moreover, if only gluon fusion production channels are considered,  the coupling $c_{GG}$ mimics the effect of $\hat{c}_H$. Therefore, to disentangle the effect of $c_{GG}$ and constrain $\hat{c}_H$, Higgs production cross-sections in different channels have to be compared; in particular the weakly sensitive vector-boson fusion (VBF) channels have to be considered. 

{
\renewcommand{\arraystretch}{1.1} 
\begin{table}[t]
\small
\begin{center}
\begin{tabular}{c|c|C{3cm} }
\textbf{Coupling} & \textbf{Direct Constraint} & \textbf{RG-induced Constraint}  \\
\hline
$\hat{c}_{S}(m_t)$ & $[-1,2]\times 10^{-3}$~\cite{gfitter} &-\\
 $\hat{c}_T(m_t)$ & $[-1,2] \times 10^{-3}$~\cite{gfitter} &- \\ 
$\hat{c}_{Y}(m_t)$ &$[-3, 3] \times 10^{-3}$~\cite{Barbieri} &- \\ 
$\hat{c}_{W}(m_t)$ &$[-2,2] \times 10^{-3}$~\cite{Barbieri}  &-\\
$\hat{c}_{\gamma \gamma}(m_t)$ &  $[-1,2] \times 10^{-3}$~\cite{Pomarol:2013zra} &-\\
$\hat{c}_{\gamma Z}(m_t)$ &$[-0.6, 1] \times 10^{-2}$~\cite{Pomarol:2013zra}&$[-2, 6] \times 10^{-2}$\\ 
$\hat{c}_{\kappa\gamma}(m_t)$ &$[-10, 7] \times 10^{-2}$~\cite{leptgc}&$[-5,  2] \times 10^{-2}$\\
$\hat{c}_{gZ}(m_t)$ &$[-4, 2] \times 10^{-2}$~\cite{leptgc}&$[-3, 1] \times 10^{-2}$\\
$\hat{c}_{\lambda\gamma}(m_t)$ &$[-6, 2] \times 10^{-2}$~\cite{leptgc}&$[-2, 8] \times 10^{-2}$\\
$\hat{c}_H(m_t)$ & $[-6, 5] \times 10^{-1}$~\cite{Francesco} &$[-2, 0.5] \times 10^{-1}$ \\ 
 \end{tabular}
\caption{\small 
In this table we present the 95 $\%$ CL, direct constraints on the coefficients in the observable basis (second column). The constraints on $\hat{S}$ and $\hat{T}$ presented here the ones obtained after marginalizing on the other parameters in the fit of Ref.~\cite{gfitter}. In the analysis we use the $\hat{S},\hat{T}$-ellipse from Ref.~\cite{gfitter} with $U=0$. Simultaneous constraints on all three of the TGC observables do not exist in the literature, so we have provided the individual constraints on the three couplings  without taking into account correlations between them~\cite{leptgc}. In the third column we show the RG-induced constraint we are able to obtain under the assumption of no fine-tuning in \eq{eq:NumericalRG}, for $\Lambda = 2\mbox{ TeV}$. }\label{table:ObsBasisNum}
\end{center}
\end{table}
}

Based on their precision of measurement,  the observables can be divided into at least two groups. In the first group, containing highly constrained operators, we have the four electroweak parameters and the Higgs diphoton coupling (see Table~\ref{table:ObsBasisNum}),
\be
\{\hat{c}_S,~\hat{c}_T,~\hat{c}_W,~\hat{c}_Y,~\hat{c}_{\gamma\gamma}\} \ ,
\label{frs}
\ee
which have been measured at the per mille level. In the second group we have the $h \gamma Z$ coupling, the couplings related to the three TGC observables $\kappa_\gamma, g^1_Z, \lambda_\gamma$ and $\hat{c}_H$,
\be
\{\hat{c}_{\gamma Z},~\hat{c}_{\kappa \gamma},~\hat{c}_{gz},~\hat{c}_{\lambda\gamma}, c_H\} \ ,
\label{sec}
\ee
which are much more weakly constrained. One can, in fact, further split the above set into $c_H$ which is constrained only at the ${\cal O}(1)$ level and the other couplings that are constrained at the few percent level.

We are interested in finding instances where the  couplings from the second group in \eq{sec} appear in the RGE's of the first group of couplings in \eq{frs}. To check this we rotate the anomalous dimension matrix to the observable basis defined by \eq{eq:ObsCoeffEWPT}, \eq{eq:ObsCoeffTGC},  and \eq{eq:ObsCoeffHiggs}. We present the anomalous dimension matrix in the observable basis in Table~\ref{tab:ObsBasisMatrix}. Using this, and fixing $\Lambda = 2$\, TeV, we write numerically \eq{ObsCoeffRunning} as
\begin{gather}
\label{eq:NumericalRG}
\left( \hat{c}_S,\hat{c}_T,\hat{c}_Y,\hat{c}_W,\hat{c}_{\gamma\gamma},\hat{c}_{\gamma Z},\hat{c}_{\kappa \gamma},\hat{c}_{gz},\hat{c}_{\lambda\gamma}, \hat{c}_H \right)^t(m_t) \simeq \\[0.2cm]
{\footnotesize \left(\!\!\! \begin{array}{cccccccccc}
 0.9 & 0.003 & -0.03 & -0.08 & -0.02 & -0.02 & -0.04 & 0.05 & -0.01 & 0.001 \\
 0.03 & 0.8 & -0.02 & -0.009 & 0 & 0 & -0.03 & 0.01 & 0 & -0.003 \\
 0.001 & 0 & 0.9 & 0 & 0 & 0 & -0.001 & 0.001 & 0 & 0 \\
 0 & 0 & -0.001 & 0.8 & 0 & 0 & 0 & -0.003 & 0 & 0 \\
 0 & 0 & 0 & 0 & 0.9 & 0 & 0.006 & 0 & 0.02 & 0 \\
 0 & 0 & 0 & 0 & 0 & 0.9 & 0.007 & 0 & 0.03 & 0 \\
 0 & 0 & 0 & 0 & -0.02 & -0.02 & 0.9 & 0 & -0.01 & 0 \\
 0.0004 & -0.0007 & -0.0004 & 0.1 & 0 & 0 & -0.0004 & 0.9 & 0 & -0.0007 \\
 0 & 0 & 0 & 0 & 0 & 0 & 0 & 0 & 0.9 & 0 \\
 -0.02 & 0.03 & 0.01 & -0.4 & 0 & 0 & 0.02 & -0.3 & 0 & 0.8 \\
\end{array} \!\!\! \right) \left(
 \!\!\!\! \begin{array}{c}
 \hat{c}_S(\Lambda) \\
 \hat{c}_T(\Lambda) \\
 \hat{c}_Y(\Lambda) \\ \hat{c}_W(\Lambda)
 \\ \hat{c}_{\gamma\gamma}(\Lambda)
 \\ \hat{c}_{\gamma Z}(\Lambda)
 \\ \hat{c}_{\kappa \gamma}(\Lambda)
 \\ \hat{c}_{gz}(\Lambda)
 \\ \hat{c}_{\lambda\gamma}(\Lambda)
 \\ \hat{c}_H(\Lambda) 
\end{array} \!\!\!\!
\right). \nonumber  
}
\end{gather}
We can now derive the RG-induced constraints by using \eq{eq:RGinducedBounds} assuming no fine-tuning among the different terms in the RGE's. 

\begin{figure}[t]
\begin{center}
\hspace*{-0.65cm} 
\begin{minipage}{0.5\linewidth}
\begin{center}
	\hspace*{0cm} 
	\includegraphics[width=8.5cm]{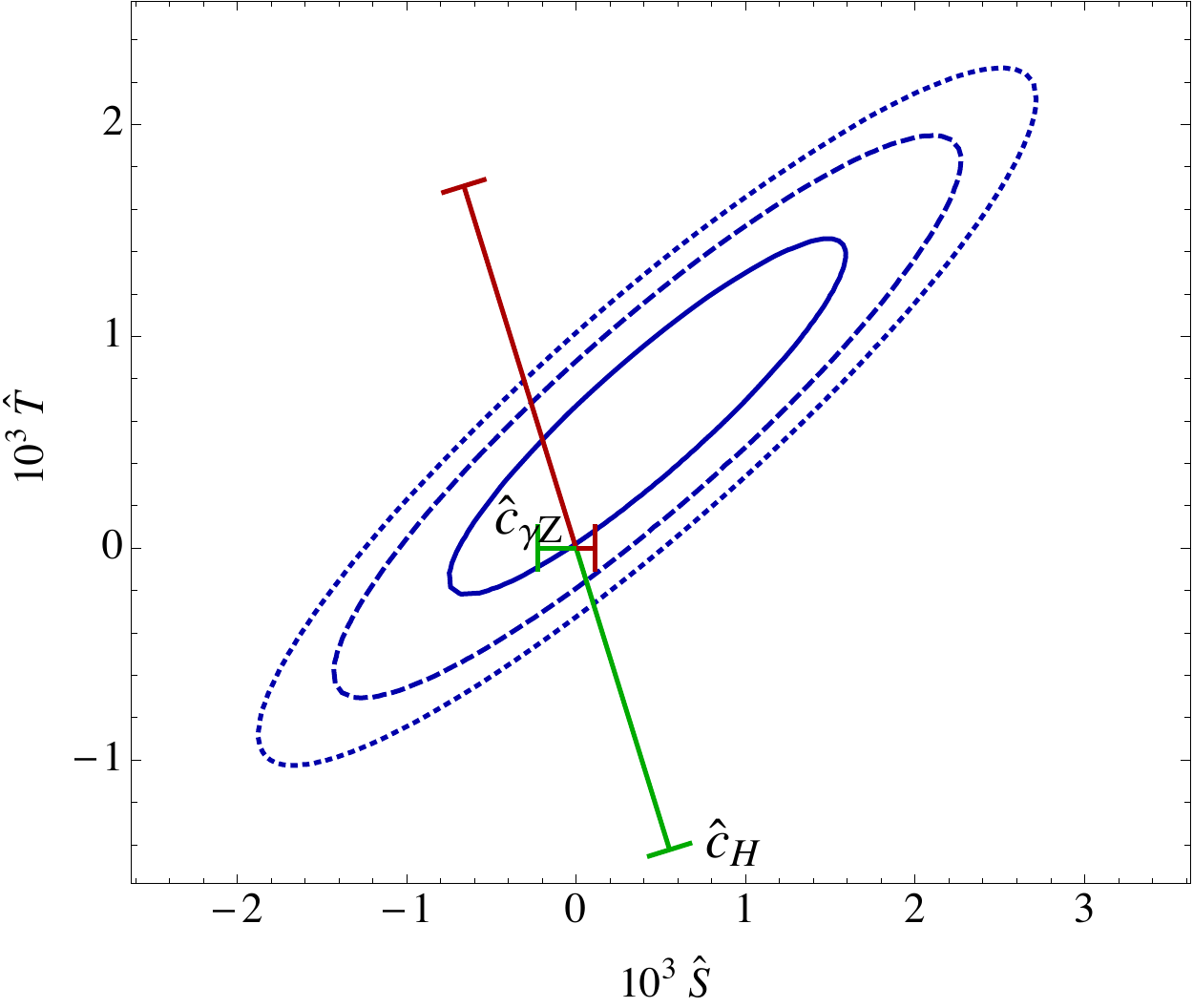}\\
	\mbox{\footnotesize (a)}
\end{center}
\end{minipage}
\hspace{0.25cm}
\begin{minipage}{0.5\linewidth}
\begin{center}
	\hspace*{0cm} 
	\includegraphics[width=8.5cm]{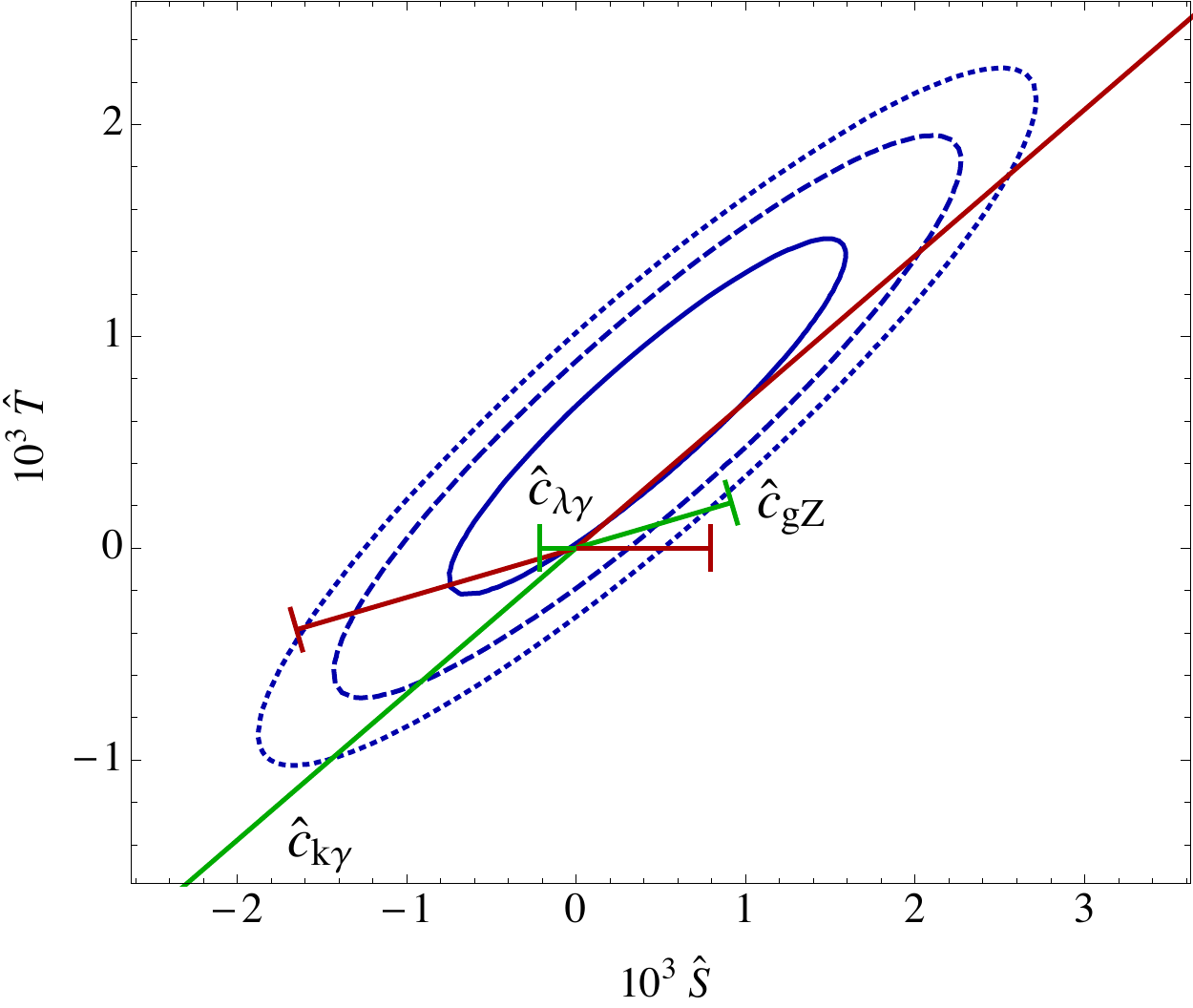}\\
	\mbox{\footnotesize (b)}
\end{center}
\end{minipage}
\end{center}
\begin{center}
\caption{\small The blue ellipses represent the 68\% (solid), 95\% (dashed) and 99\% (dotted) CL bounds on $\hat{S}$ and $\hat{T}$ as obtained in the fit of Ref.~\cite{gfitter} with $U = 0$. The straight lines represent the RG-induced contribution to the oblique parameters from the weakly constrained observable couplings of \eq{sec}, divided in Higgs couplings (a) and TGC couplings (b), using the first two lines of \eq{eq:NumericalRG}, for $\Lambda = 2 \mbox{ TeV}$. The length of the lines corresponds to their present 95\% CL direct bounds, see Table \ref{table:ObsBasisNum}; the line is green (red) for positive (negative) values of the parameters. \label{fig:RGBounds}}
\end{center}
\end{figure}

The strongest RG-induced constraints come from the direct bounds on the $\hat{S}, \hat{T}, W$ and $Y$ parameters, i.e. the first four lines in \eq{eq:NumericalRG}. We require  that each  observable coupling individually satisfies the four RG-induced constraints from these electroweak precision parameters simultaneously. It is very important to take into account the experimental correlations between $\hat{S}, \hat{T}, W$ and $Y$ while imposing these bounds\cite{Ciuchini:2013pca, Grojean:2013qca, Contino:2013gna}. Note that the RG-mixing contributions to $\hat{c}_W$ and $\hat{c}_Y$, from  the couplings in the weakly constrained group in \eq{sec}, is either absent or accidentally much smaller than the ones  to $\hat c_S$ and $\hat c_T$ (see the RG contributions to $\hat{c}_W$ and $\hat{c}_Y$ in the third and fourth row of \eq{eq:NumericalRG}). We, therefore,  look at the constraints on the $\hat{S}-\hat{T}$ plane taking $W=Y=0$. We use  the $\hat{S}-\hat{T}$ ellipse in ref.~\cite{gfitter}, which assumes  $W=Y=U=0$, to derive our constraints. We present these RG-induced bounds and compare them with the direct bounds  in Table~\ref{table:ObsBasisNum} and in Figure~\ref{fig:RGBounds}. We find that for each of the couplings in the second group we can derive a  RG-induced constraint  stronger  than, or of the same order of,  the direct tree-level constraint.  We also obtain RG-induced bounds from the direct constraint on $\hat{c}_{\gamma \gamma}$  using the fifth line in \eq{eq:NumericalRG} and \eq{eq:RGinducedBounds},
\be \begin{split}
	\hat{c}_{\kappa \gamma} &\in [-0.2, 0.3] \ ,\\
	\hat{c}_{\lambda \gamma} &\in [-0.05, 0.10] \ ,
\end{split} \ee
but  at present these bounds are weaker than those from the direct bounds on electroweak parameters.

Let us briefly comment on  alternate choices for our observable basis. In general, a change of \emph{observable basis} modifies the anomalous dimension matrix of Table~\ref{tab:ObsBasisMatrix}, also for the observables which were maintained in the basis. Thus, the RG-induced constraints we have derived, are applicable only to our particular choice of observables, and for an alternate choice the analysis must be repeated.\footnote{Note that for our choice of observable basis, $h\to \gamma\gamma$ does not receive a contribution from the $\hat S$ parameter even though there is a dependance  on $c_{WB}$ in the anomalous dimension but $c_{WB}$ is actually reconstructing the $\delta \kappa_{\gamma}$ parameter.}
For instance, the Higgs decay observables related to $h\to W^+ W^-, ZZ$ decays could have been alternatively chosen as part of our observable basis instead of two of the TGC observables ($\kappa_\gamma$ and $g_Z$) but we have kept the TGC in our basis as they are measured more precisely than these Higgs decay observables. This situation is likely to continue in the future. Although, observables like the relative deviation of $h\to W^+ W^-, ZZ$ with respect to the SM would  be strongly constrained at the 5 $\%$(3 $\%$) level at the LHC with 300\,fb$^{-1}$ ( 3000\,fb$^{-1}$) data~\cite{Dawson:2013bba}, the bounds on TGC are also expected to become stronger by an order of magnitude at the LHC~\cite{Dawson:2013bba} so that the TGC would still be more precisely measured than these Higgs observables. At linear colliders the Higgs $h\to W^+ W^-, ZZ$ is expected to be measured at the level of 0.5 $\%$~\cite{Dawson:2013bba} and the TGC observables at the  $10^{-4}$ level  \cite{ILC}; again the TGC observables would be more constrained.

{
\renewcommand{\arraystretch}{2.} 
\renewcommand{\tabcolsep}{0.2mm}

\begin{table}[tdp]

\begin{center}
\scriptsize
\begin{tabular}{c|ccccc}
 & $\hat{c}_{S}$  & $\hat{c}_{T}$  & $\hat{c}_{Y}$  & $\hat{c}_{W}$  & $\hat{c}_{\gamma\gamma}$ \tabularnewline
\hline 
$\gamma_{\hat{c}_{S}}$  & $\frac{1}{3}g^{\prime2}+6y_{t}^2$  & $-\frac{g^{2}}{2}$  & $\frac{1}{8}g^{\prime2}\left(147-106\frac{g^{\prime2}}{g^{2}}\right)$  & $\frac{1}{8} \left(77 g^{2} + 58g^{\prime2} \right)$  & $16e^{2}$ \tabularnewline
$\gamma_{\hat{c}_{T}}$  & $-9 g^{\prime2}-24t_{\theta_W}^2 \lambda$  & $\frac{9}{2}g^{2} + 12 y_{t}^2+12 \lambda$  & $\frac{9}{2}g^{\prime 2} + 12 t_{\theta_W}^2 (g^{\prime 2} + \lambda)$  & $\frac{9}{2} g^{\prime2}$  & $0$ \tabularnewline
$\gamma_{\hat{c}_{Y}}$  & $-\frac{2}{3}g^{\prime2}$  & $0$  & $\frac{94}{3}g^{\prime2}$  & $0$  & $0$ \tabularnewline
$\gamma_{\hat{c}_{W}}$  & $0$  & $0$  & $\frac{53}{12}g^{\prime2}\left(1-3 t_{\theta_W}^2\right)$  & $\frac{331}{12}g^{2}+\frac{29}{4}g^{\prime2}$  & $0$ \tabularnewline
$\gamma_{\hat{c}_{\gamma\gamma}}$  & $0$  & $0$  & $0$  & $0$  & $-\frac{9}{2}g^{2}-\frac{3}{2}g^{\prime2}+6y_{t}^2+12\lambda$ \tabularnewline
$\gamma_{\hat{c}_{H}}$  & $18 g^{\prime2} -t_{\theta_W}^2( 9 g^{\prime 2} + 24 \lambda )$  & $~~-9g^{2}+\frac{9}{2}g^{\prime2}+12\lambda$  & $~~t_{\theta_W}^2 \left( - \frac{141}{4} g^{\prime 2} + 12 \lambda \right)$  & $~~\frac{63}{2}g^2 + \frac{51}{4} g^{\prime 2} + 72 \lambda$  & $0$ \tabularnewline
$\gamma_{\hat{c}_{\gamma Z}}$  & $0$  & $0$  & $0$  & $0$  & $0$\tabularnewline
$\gamma_{\hat{c}_{kZ}}$  & $0$  & $0$  & $0$  & $0$  & $-16e^{2}$\tabularnewline
$\gamma_{\hat{c}_{gZ}}$  & $-\frac{g^{\prime2}}{6c_{\theta_W}^{2}}$  & $\frac{g^2}{12 c_{\theta_W}^2}$  & $\frac{g^{\prime2}}{8 c_{\theta_W}^2} (106 t_{\theta_W}^2 - 29)$  & $-\frac{1}{8c_{\theta_W}^{2}}(79g^{2}+58g^{\prime2})$  & $0$ \tabularnewline
$\gamma_{\hat{c}_{\lambda\gamma}}$  & $0$  & $0$  & $0$  & $0$  & $0$ \tabularnewline
\end{tabular}
\end{center}
\vspace{1cm}
\hspace{-1.3cm}
\scriptsize
\begin{tabular}{c|ccccc}
 & $\hat{c}_{H}$  & $\hat{c}_{\gamma Z}$  & $\hat{c}_{\kappa \gamma}$  & $\hat{c}_{gZ}$  & $\hat{c}_{\lambda\gamma}$ \tabularnewline
\hline 
$ $$\gamma_{\hat{c}_{S}}$  & $-\frac{1}{6}g^{2}$  & $4(g^{2}-g^{\prime2})$  & $-\frac{11}{2}g^{2}-\frac{1}{6}g^{\prime2} - 4\lambda$  & $c_{\theta_W}^{2}\left(9 g^{2} - \frac{1}{3} g^{\prime2} \right)$  & $-2g^{2}$ \tabularnewline
$\gamma_{\hat{c}_{T}}$  & $\frac{3}{2}g^{\prime2}$  & $0$  & $-9 g^{\prime 2} - 24 t_{\theta_W}^2 \lambda$  & $24 s_{\theta_W}^2 \lambda$  & $0$ \tabularnewline
$\gamma_{\hat{c}_{Y}}$  & $0$  & $0$  & $-\frac{2}{3}g^{\prime 2}$  & $\frac{2}{3}e^{2}$  & $0$ \tabularnewline
$\gamma_{\hat{c}_{W}}$  & $0$  & $0$  & $0$  & $- \frac{2}{3}c_{\theta_W}^{2}g^{2}$  & $0$ \tabularnewline
$\gamma_{\hat{c}_{\gamma\gamma}}$  & $0$  & $0$  & $\frac{3}{2}g^{2}-2\lambda$  & $0$  & $3g^{2}$ \tabularnewline
$\gamma_{\hat{c}_{H}}$  & $-\frac{9}{2}g^{2}-3g^{\prime2}+12y_{t}^2+24\lambda$  & $0$  & $9 g^{\prime 2} (2 - t_{\theta_W}^2) - 24 t_{\theta_W}^2 \lambda$  & $9(g^{\prime 2} s_{\theta_W}^2 - g^2 c_{\theta_W}^2) - 24 \lambda (6 c_{\theta_W}^2 - s_{\theta_W}^2)$  & $0$ \tabularnewline
$\gamma_{\hat{c}_{\gamma Z}}$  & $0$  & $-\frac{7}{2}g^{2}-\frac{1}{2}g^{\prime2}+6y_{t}^2+12\lambda$  & $~~c_{\theta_W}^{2}(2g^{2}-2\lambda)-s_{\theta_W}^{2}(g^{2}-2\lambda)$  & $0$  & $\frac{g^{2}}{2}(11c_{\theta_W}^{2}-s_{\theta_W}^{2})$ \tabularnewline
$\gamma_{\hat{c}_{\kappa\gamma}}$  & $0$  & $4(g^{2}-g^{\prime2})$  & $\frac{11}{2}g^{2}+\frac{g^{\prime2}}{2}+6y_{t}^2+4\lambda$  & $0$  & $2g^{2}$ \tabularnewline
$\gamma_{\hat{c}_{gZ}}$  & $\frac{g^2}{12 c_{\theta_W}^2}$  & $0$  & $\frac{g^{\prime2}}{6c_{\theta_W}^2}$  & $\frac{17}{2}g^{2}-\frac{g^{\prime2}}{6}+6y_{t}^2$  & $0$ \tabularnewline
$\gamma_{\hat{c}_{\lambda\gamma}}$  & $0$  & $0$  & $0$  & $0$  & $\frac{53}{3} g^2$ \tabularnewline
\end{tabular}

\caption{\small Anomalous dimension matrix in the observables basis. We defined $t_{\theta_W} = \tan \theta_W$.}
\label{tab:ObsBasisMatrix}

\end{table}
}

{\renewcommand{\arraystretch}{1.1} 
\begin{table}[t]
\small
\begin{center}
\begin{tabular}{c|c|c|c|c|c|c}
\textbf{Direct } & \textbf{Future} &$|\hat{c}_{\kappa \gamma}|$ &$|\hat{c}_{\gamma Z}|$ & $|\hat{c}_{\lambda \gamma}|$& $|\hat{c}_{H}|$ \tabularnewline
\textbf{Measurement}  & \textbf{Precision} & && & & \tabularnewline\hline
$\hat{c}_{\gamma \gamma}$& $4\times 10^{-5}$~\cite{Dawson:2013bba}& $6 \times 10^{-3}$&-&$2 \times 10^{-3}$&- \tabularnewline
$\hat{c}_{\gamma Z}$&$3\times 10^{-4}$~\cite{Dawson:2013bba}&$4 \times 10^{-2}$&-&$1 \times 10^{-2}$&- \tabularnewline
$\hat{c}_{\kappa \gamma}$&$2\times 10^{-4}$~\cite{ILC}&- &$1 \times 10^{-2}$&$1 \times 10^{-2}$&- \tabularnewline
$\hat{c}_{gZ}$&$2\times 10^{-4}$~\cite{ILC} &0.4 &-&-&0.25\tabularnewline
\end{tabular}
\end{center}
\caption{\small In this table we present the minimum value of the couplings in \eq{sec} to which direct measurements of the observables in the first column would be sensitive via the one loop RG-mixing effects computed in this work. The long term projection for the measurement precision for the observables in the first column is given in the second column.}
\label{future}
\end{table}
}

Finally, let us discuss the future prospects for these RG-induced effects. In the future, as the measurement  of the observables we have considered becomes more and more precise, it may be possible to detect signs of new physics. In this case, since some of the observables in Table~\ref{table:ObsBasisNum} will be non-zero one would expect a deviation, via RG-mixing, also in other observables, unrelated at tree level.
Note that according to future projections, $\hat{c}_{\gamma \gamma}$, the TGC observables ($\hat{c}_{\kappa \gamma},~\hat{c}_{gz}$) and $\hat{c}_{\gamma Z}$ would  be measured at the $10^{-4}$ level~\cite{Dawson:2013bba, ILC} at linear colliders and thus all these observables would be sensitive to RG-induced mixing effects of the couplings in \eq{sec}, if they are above a minimal value.\,\footnote{Future prospects for measurements at the $Z$-pole predict an enhancement of the precision, with respect to the present one, of about one order of magnitude for ILC \cite{ILC} and two orders of magnitude for TLEP \cite{Gomez-Ceballos:2013zzn}, depending on the observable. Moreover, from runs at energy $\sqrt{s}\sim 2 m_W$, the measurement of the $W$ mass is predicted to became more precise by one (ILC) or two (TLEP) orders of magnitude. This will imply an enhancement of the precision in the oblique parameters $\hat{S}$, $\hat{T}$, $W$ and $Y$. A more detailed study of these future prospects is beyond the scope of this paper, since our aim is only to show some examples for future applications of the general idea of RG-induced bounds.} We present these minimum values in Table~\ref{future}. If, instead, a deviation is detected in some observable but no such RG-induced deviation in other observables is detected at the level hinted by our analysis, then this would indicate a tuning (or a correlation) among the various RG contributions to the direct measurement, see \eq{eq:tuning}. Take, for example, the first row of Table~\ref{future}. Suppose we measure the deviation  $\hat{c}_{\lambda \gamma} \sim 1 \times 10^{-2}$, a value larger than the minimum value presented in Table  \ref{future}, while instead $h \rightarrow \gamma\gamma$ would still remain compatible with zero with the reported sensitivity. From \eq{eq:tuning} we would than conclude that a fine-tuning of the order $\Delta_{\gamma\gamma} \gtrsim 5$ would be necessary to accommodate the data, or that some particular correlation in the UV physics is needed to induce such cancellation.

\section{Scaling of the gluon operators}
\label{GluonSect}

In this section we shall extend the results of the previous sections and present also the scaling of the bosonic operators that contain gluons, as defined in Table~\ref{table:one1}:
\be
	\{ \Op_{2G}, \ \Op_{GG}, \ \Op_{3G}\}.
\ee
The anomalous dimension matrix is shown in Table~\ref{tab:OurBasisMatrixGluons}, where the $c_{3G}$ self-renormalization has been taken from refs.~\cite{Narison:1983,Morozov:1985ef}. This matrix already contains the effect of the redundant operators that are generated radiatively and, upon eliminating them,  modify the RG of the operators in Table~\ref{table:one1}, see Appendix~\ref{appen} for details.
 
In the same spirit of Section~\ref{Sconst}, let us now turn to the observables which are sensitive to these operators and review the present constraints.
The Wilson coefficient $c_{2G}$ can be put in one-to-one relation to the parameter $Z$ introduced in ref.~\cite{Barbieri} (analogous to the $W$ and $Y$ electroweak parameters):
\be
	Z = \frac{m_W^2}{\Lambda^2} c_{2G}.
\ee
A bound on this parameter has been obtained by an analysis of dijets events at LHC~\cite{Domenech:2012ai}:
\be
	-9 \times 10^{-4} \lesssim Z
	\lesssim 3 \times 10^{-4}.
\ee
A bound on $c_{GG}$ can be obtained from the analysis of the Higgs production cross section at LHC. The relevant phenomenological Lagrangian is
\be
	\lag_h \supset \hat{c}_{GG} \frac{h v}{m_W^2} g_s^2 G^A_{\mu\nu} G^{\mu\nu \, A},
\ee
where we defined
\be
	\hat{c}_{GG} \equiv \frac{m_W^2}{\Lambda^2} c_{GG}.
\ee
The most recent bound, obtained in ref.~\cite{Pomarol:2013zra} after marginalizing over the other deviations from the SM, reads
\be
	\hat{c}_{GG} \in [-0.8, 0.8] \times 10^{-3}.
\ee
The coefficient $c_{3G}$, analogous to the $SU(2)_L$ counterpart $c_{3W}$, would contribute to the anomalous triple gluon couplings. These effects can be measured at LEP, Tevatron and LHC, for example via top-quark pair production, see for example ref.~\cite{Simmons:1995hb} where it is estimated that LHC should be able to put a bound
$|\hat{c}_{3G}| \equiv \left|  c_{3G}  \right| m_W^2/\Lambda^2\lesssim 0.1.$

{
\renewcommand{\arraystretch}{2.05} 
\renewcommand{\tabcolsep}{1.3mm}

\begin{table}[tdp]
\begin{center}
\footnotesize
\begin{tabular}{c|ccccc}
 & $c_{2G}$  & $c_{GG}$ & $c_{3G}$ & $c_{2B}$ & $c_{2W}$\\
\hline 
$ $$\gamma_{c_{2G}}$  & $\frac{266}{9} g_{s}^{2}$  & $0$  & $0$  & $g^{\prime 2} \left(\frac{17}{6}(Y_u^2+Y_d^2) + 12 Y_u Y_d\right)$  & $0$ \\
$\gamma_{c_{GG}}$ & $0$  & $-\frac{3}{2}g^{\prime2}-\frac{9}{2}g^{2}+12\lambda+6y_{t}^{2}$  & $0$  & $0$  & $0$ \\
$\gamma_{c_{3G}}$ & $0$  & $0$  & $22 g_s^2$ & $0$  & $0$ 
\end{tabular}
\caption{\small Anomalous dimension matrix for the Wilson coefficients of the dim-6 bosonic operators with gluons, in the basis defined in Section~\ref{Dim6Basis}. The contributions to and from the other coefficients of the operators in \eq{univBasis}, not reported here, are zero.
\label{tab:OurBasisMatrixGluons}}
\end{center}
\end{table}
}

As can be seen in Table~\ref{tab:OurBasisMatrixGluons}, no mixing to (or from) these gluon operators is present among the operators we considered in Table~\ref{table:one1}, the only exception being a contribution from $c_{2B}$ to $c_{2G}$ which, however, is not very interesting since $c_{2B}$ is already very well directly constrained by the oblique $Y$ parameter.
For this reason, we are not able to cast any indirect constraint using these gluon operators.

\section{Conclusions}
\label{Sec:Conclusions}

We computed the scaling and mixing of 13 dim-6 deformations of the SM affecting EW precision observables (4), anomalous EW triple gauge boson couplings (3), QCD observables (2) and Higgs decays (4). This computation has  important phenomenological implications.
Particularly interesting is the RG-mixing induced among 10 of these observables (the 2 two QCD observables and one Higgs observable, namely $\Gamma (h \to gg)$, constitute a separate sector that does not mix in a relevant way with the severely constrained EW observables.).

These 10 different observables are constrained at very different levels of precision. For example, whereas the electroweak precision observables and the operator coefficient related to the $h\to \gamma\gamma$ partial width are constrained at the per mille level, the TGC and  the 2 other Higgs observables are constrained at the percent level at most.  
As we run down from the new physics scale  to the lower scale of experiments, quantum effects mix the observables
and the most severely constrained ones receive a contribution from the ones allowed to deviate the most from the SM predictions.
These RG-contributions could in principle be of the same size or even larger than the direct experimental bounds, in other words, the difference in the experimental sensitivities can compensate for the RG-loop factor. Requiring that these RG-contributions do obey individually the direct bounds, i.e. dismissing any possible tuning/correlation among the various RG-terms, we can derive some indirect RG-induced bounds on the weakly constrained observables from the direct measurement of the severely constrained  ones. This analysis is particularly relevant for the TGC and the universal shift of the Higgs couplings, as reported in Table~\ref{table:ObsBasisNum}.

We also looked at the future prospects of these RG-induced effects. If a deviation from the SM is observed in some of the observables we considered, in the absence of tuning one would expect a deviation, due to these RG effects, to appear also in other seemingly unrelated observables.  If, instead, these RG-induced deviations are not observed, it would mean that some tuning is needed, or it would indicate some correlation among the higher dimensional operators pointing towards a particular structure of the new physics that has been integrated out. We have presented the projected future experimental sensitivity to these RG effects in Table~\ref{future}.

The first run of the LHC ended beautifully with the discovery of the Higgs boson and initiated an era of measurements in the EWSB sector that remained only indirectly constrained for several decades. With the next run of the LHC and the high-luminosity program  will start an era of precision that will lead certainly to a better understanding of  the Higgs sector itself and  also, hopefully, to the first glimpse of the new physics laying beyond the Standard Model. We hope that the results we presented in this paper will be a powerful tool in that quest.

\vspace{.2cm}
\noindent {\bf Note added:} While this paper was being submitted, the work~\cite{AJMT} appeared. It computed the gauge-coupling dependence of the anomalous dimensions among the dim-6 operators. 

\section*{Acknowledgments}
We thank  J.R.~Espinosa, E.~Masso, A.~Pineda, A.~Pomarol and F.~Riva for insightful discussions and comments.
The work of J.E.M.~has been supported by the Spanish Ministry MECD through the FPU grant AP2010-3193.
C.G.~is  supported by the Spanish Ministry MICINN under contract
FPA2010-17747 and by the European Commission under the ERC Advanced Grant 226371 \emph{MassTeV} and the contract PITN-GA-2009-237920 \emph{UNILHC}.
D.M.~thanks the Theory Division at CERN for hospitality and the \emph{UNILHC} network for support during the completion of this work. 

\appendix

\section{Dealing with redundant operators} 
\label{appen}

In this appendix we explain in detail the anomalous dimension matrix presented in the main body of the paper, Tables~\ref{tab:OurBasisMatrix} and~\ref{tab:OurBasisMatrixGluons}. As remarked in Section~\ref{Dim6Basis}, a common effect encountered in the computation of the scaling of the dim-6 operators is the appearance of counter-terms that correspond to operators not included in our basis, i.e. operators that are redundant for the description of physical processes. In particular, the set of 13 operators we are interested in,
\begin{equation}
\label{UnivBasis1}
\left\{{\cal{O}}_H, {\cal{O}}_T, {\cal{O}}_B, {\cal{O}}_W, {\cal{O}}_{2B}, {\cal{O}}_{2W}, {\cal{O}}_{BB}, {\cal{O}}_{WW}, {\cal{O}}_{WB}, {\cal{O}}_{3W}, \Op_{2G}, \Op_{GG}, \Op_{3G} \right\} \ ,
\end{equation}
not only mix among themselves under the RG flow but also generate redundant operators that are not included in our basis (defined in Section~\ref{Dim6Basis}). In this appendix we first give a pedagogic example of radiatively generated redundant operators, Section~\ref{appen0}. Then, we present the set of redundant operators generated by those in \eq{UnivBasis1}, together with their anomalous dimensions, Section~\ref{appenA1}. In Section~\ref{redred} we explain how the redundant operators are redefined back into our basis and what is their effect on the anomalous dimensions of the operator set in \eq{UnivBasis1}~\cite{EEMP2}.


\subsection{Example of radiatively generated redundant operators}
\label{appen0}

\begin{figure}[t]
\begin{center}
\hspace*{-0.65cm} 
\begin{minipage}{0.5\linewidth}
\begin{center}
	\hspace*{0cm} 
	\includegraphics[width=5cm]{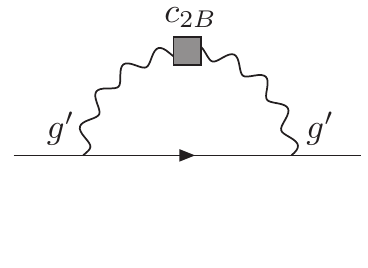}\\
	\mbox{\footnotesize (a)} \\
\end{center}
\end{minipage}
\hspace{0.25cm}
\begin{minipage}{0.5\linewidth}
\begin{center}
	\hspace*{0cm} 
	\includegraphics[width=5cm]{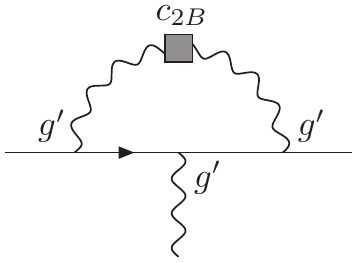}\\
	\mbox{\footnotesize (b)} \\
\end{center}
\end{minipage}
\end{center}
\begin{center}
\parbox{14cm}{
\caption{\small  Feynman diagrams representing the contribution of the dim-6 operator ${\cal O}_{2B}=-\frac{1}{2} (\partial^\mu B_{\nu\mu})^2$ to the renormalization of the vertices $ \bar{e}_R-e_R$ (diagram (a)) and $ \bar{e}_R- e_R -B^\nu$ (diagram (b)). \label{FeynGraphs_eeB}}}
\end{center}
\end{figure}

As a first step, let us give a detailed example of the generation of redundant operators by the ones in \eq{UnivBasis1}.
Consider the renormalization of the vertex $\bar{e}_R-e_R-B^{\nu}$ by the operator ${\cal O}_{2B}$. There is only one possible diagram, depicted in Fig.~\ref{FeynGraphs_eeB}(b), which can give contributions to any of the operators
\begin{equation}
\label{eq:ex1}
{\cal O}_{BR}^e = g^\prime \partial^\nu  B_{\mu\nu} (\bar{e}_R \gamma^\mu e_R )\, , \quad {\cal O}_{BR}^{\prime e}=g^{\prime} \widetilde{B}_{\mu \nu} i  \bar{f}_{R}^i  \gamma^{\mu}  D^{\nu} f_{R}^i \, ,
\end{equation}
or to the three-point vertex of the operator
\begin{equation}
\label{eq:ex2}
{\cal O}_{K3R}^e =\frac{1}{2} \bar{e}_R (\Dslash D^2+D^2\Dslash)e_R \, . 
\end{equation}
It can be easily checked that there is no other operator with the same field content which is also independent from the ones in \eq{eq:ex1} and \eq{eq:ex2}. As for a CP-odd version of ${\cal O}_{BR}^{\prime e}$ is of not concern to us since it is clear that the diagrams we are considering cannot violate CP. 
The crucial point of this discussion is that the above operators are not contained in our basis, therefore one has to redefine them back to the ones in our basis, giving a contribution in the anomalous dimensions. These indirect contributions of ${\cal O}_{2B}$ to the anomalous dimensions of the bosonic operators are of the same order as the direct contributions computed via one-particle-irreducible diagrams, 
it is therefore necessary to keep track of all such effects in order to have a consistent calculation.

The computation of diagram (b) in Fig.~\ref{FeynGraphs_eeB} gives us, in general, a combination of the contributions from $\Op_{2B}$ to all the operators in \eq{eq:ex1} and \eq{eq:ex2}. 
To disentangle the different contributions from the divergent part of diagram (b), we look at the different momentum structures of the three operators and recognize them inside the result of diagram (b):
\be
	\mathcal{A}_{(b)}^\text{div} = - \frac{\gamma_{c_{K3R} | c_{2B}}^{(b)} }{2 \epsilon} c_{2B} \langle \Op_{K3R}^e \rangle - \frac{\gamma_{c_{BR} | c_{2B}}}{2 \epsilon} c_{2B} \langle \Op_{BR}^e \rangle - \frac{\gamma_{c_{BR}^\prime | c_{2B}}}{2 \epsilon} c_{2B} \langle \Op_{BR}^{\prime e} \rangle \, ,
\ee
using dimensional regularization with $D=4-2\epsilon$. $\langle \Op_{i} \rangle$ represents the Feynman rule of the operator $\Op_{i}$ with the external states of diagram (b). We obtain
\be
	16 \pi^2 \gamma_{c_{K3R} | c_{2B}}^{(b)} = - Y_e^2 g^{\prime 2}, \quad
	16 \pi^2 \gamma_{c_{BR} | c_{2B}} = - \frac{5}{6} Y_e^2 g^{\prime 2}, \quad
	16 \pi^2 \gamma_{c_{BR}^\prime | c_{2B}} = - Y_e^2 g^{\prime 2}.
	\label{comp1}
\ee

Diagram (a) in Fig.~\ref{FeynGraphs_eeB} gives univocally the contribution of ${\cal O}_{2B}$ to ${\cal O}_{K3R}^e$, since this is the only dim-6 operator with only $\{\bar{e}_R, e_R\}$ as external legs.
Since we are working in the background field gauge, we expect that all vertices of a gauge invariant operator should be renormalized in the same way.\footnote{This is somewhat trivial for this example since the considered diagrams are clearly independent of the background field gauge terms, \eq{bkgFG}; but it is relevant in general.} For this reason, and from the previous computation in \eq{comp1}, we already know what that the result of diagram (a) should be: $ \gamma_{K3R | 2B}^{(a)} =  \gamma_{K3R | 2B}^{(b)}$; and indeed this is what we find performing the explicit calculation. One can often use similar relations as a check of the computation.

In the following section we shall study how these redundant operators, generated by the RG flow, can be redefined into the ones of our basis. For instance, we shall see that the contribution from $\Op_{2B}$ to $\Op_{BR}^e$ described above has to be included as a contribution to the running of $\Op_{B}$ and $\Op_{2B}$, \eq{eq:PhysicalWilsComb}.

\subsection{Anomalous dimension matrix}
\label{appenA1}

The relevant redundant operators that are radiatively generated by those in \eq{UnivBasis1} are:
\be \begin{array}{l l}
 {\cal O}_{r} = \left| D H \right|^2 \left|H\right|^2 \, , &   {\cal O}_{K4}= \left| D^2 H \right|^2  \, , \\[0.2cm]
   {\cal O}_{LL}^{(3)}= (\bar{L}_L\sigma^a \gamma^\mu L_L)(\bar{L}_L\sigma^a \gamma^\mu L_L) \, , &  {\cal O}_L^{(3)L_1}=   i( H^\dagger  \sigma^a{\lra { D_\mu}} H)  \bar{L}_{L}^1 \sigma^a \gamma^{\mu} L_{L}^1 \, ,   \\[0.2cm]
   \Op_{RR}^{(8)u_1 d_1}=(\bar{u}_R \gamma^\mu T^A u_R) (\bar{d}_R \gamma^\mu T^A d_R)  \, , &  {\cal O}_{RR}^{e_1}=(\bar{e}_R \gamma^\mu e_R) (\bar{e}_R \gamma^\mu e_R) \, , \\[0.2cm]
 {\cal O}_{K3L}^{F_i}= \frac{1}{2}  \bar{F}_L^i \left( \Dslash D^2+  D^2 \Dslash  \right) F_L^i \, , &    {\cal O}_{K3R}^{f_i}= \frac{1}{2}  \bar{f}_R^i   \left( \Dslash D^2+  D^2 \Dslash  \right)  f_R^i \, , \\[0.2cm]
{\cal O}_{WL}^{F_i}=  g D^{\nu} W^a_{\mu \nu} (\bar{F}_{L}^i \sigma^a \gamma^{\mu} F_{L}^i ) \, , & {\cal O}_{WL}^{\prime F_i}= g \widetilde{W}_{\mu \nu}^a i \bar{F}_{L}^i \sigma^a \gamma^{\mu}  D^{\nu} F_{L}^i  \, ,   \\[0.2cm]
 {\cal O}_{BL}^{F_i}=  g^{\prime} D^{\nu} B_{\mu \nu} ( \bar{F}_{L}^i  \gamma^{\mu} F_{L}^i ) \, , & {\cal O}_{BL}^{\prime F_i}= g^{\prime} \widetilde{B}_{\mu \nu} i \bar{F}_{L}^i  \gamma^{\mu}  D^{\nu} F_{L}^i  \, ,   \\[0.2cm]
 {\cal O}_{BR}^{f_i}= g^{\prime} D^{\nu} B_{\mu \nu} ( \bar{f}_{R}^i  \gamma^{\mu} f_{R}^i ) \, , & {\cal O}_{BR}^{\prime f_i}= g^{\prime} \widetilde{B}_{\mu \nu} i  \bar{f}_{R}^i  \gamma^{\mu}  D^{\nu} f_{R}^i  \, ,   \\[0.2cm]
 {\cal O}_{GL}^{Q_i}=  g_s D^{\nu} G^A_{\mu \nu} (\bar{Q}_{L}^i T^A \gamma^{\mu} Q_{L}^i )  \, , &   {\cal O}_{GL}^{\prime Q_i}=  g_s \widetilde{G}^A_{\mu \nu} i (\bar{Q}_{L}^i T^A \gamma^{\mu} D^\nu Q_{L}^i ) \, ,   \\[0.2cm]
  {\cal O}_{GR}^{q_i}=  g_s D^{\nu} G^A_{\mu \nu} (\bar{q}_{R}^i T^A \gamma^{\mu} q_{R}^i ) \, , &  {\cal O}_{GR}^{\prime q_i}=  g_s \widetilde{G}^A_{\mu \nu} i (\bar{q}_{R}^i T^A \gamma^{\mu} D^\nu q_{R}^i )  \, , 
\end{array}\label{radred}
\ee
 
By \emph{relevant} we mean those radiatively generated redundant operators that modify the Wilson coefficient of the operators in \eq{UnivBasis1} when the former operators are redefined into operators in our basis, defined in Section~\ref{Dim6Basis}.
 
Below we present in three different tables the anomalous dimension matrix of the operators in \eq{UnivBasis1} as well as the relevant redundant operators generated by them, \eq{radred}, at the order stated in \eq{orderCalc}. We work with arbitrary $\xi$ in the background field gauge (see \eq{bkgFG}) and use dimensional regularization. All the contributions given in Tables~\ref{tabCHRT}, \ref{tabLoop} and~\ref{tabCBW2} below arise from one-particle-irreducible Feynman diagrams, i.e. it is the one-loop renormalization of the Effective Action. 


 {
\renewcommand{\arraystretch}{2.1} 
\renewcommand{\tabcolsep}{2mm}
\begin{table}[tdp]
\begin{center}
\footnotesize
\begin{tabular}{c | c c c}
		& $c_H$ & $c_r$ &$c_T$ \\ \hline
	 $\gamma_{c_H}$       & $28\lambda +12y_t^2-3\left(\frac{5}{2}g^2+g^{\prime 2}\right)$ & $\frac{3}{2}\left(2g^2+g^{\prime 2}\right)-4\lambda$ &$8\lambda-6g^2-\frac{3}{2}g^{\prime 2}$    \\ 
	 $\gamma_{c_T}$        & $\frac{3}{2} g^{\prime 2}$  & $-\frac{3}{2}g^{\prime 2}$   & $12\lambda +12y_t^2+\frac{9}{2}g^2$ \\ 
	$\gamma_{c_B}$        & $-\frac{1}{3}$ &  $\frac{1}{3}$   &$-\frac{5}{3}$   \\ 
	 $\gamma_{c_W}$       & $-\frac{1}{3}$ &  $\frac{1}{3}$   &  $-\frac{1}{3}$\\  \hdashline
	 $\gamma_{c_r}$          & $4\lambda-3g^2$ & $20\lambda +12y_t^2-\frac{3}{2}\left(g^2+g^{\prime 2}\right)$  &$-4\lambda+3g^2-6 g^{\prime 2}$          \\ 
\end{tabular}
 
\caption{\small Anomalous dimension matrix. Further contributions of ${\cal O}_H$, ${\cal O}_r$ and ${\cal O}_T$ to other operators in \eq{UnivBasis1} and \eq{radred} are either zero or proportional to the Yukawa coupling of any fermion lighter than the top. The dashed line separates the anomalous dimension of the operators in our basis from that of the redundant operators.  }
\label{tabCHRT}
\end{center}
\end{table}
}


{
\renewcommand{\arraystretch}{1.8} 
\renewcommand{\tabcolsep}{2mm}
\begin{table}[htdp]
\scriptsize
\begin{center}
\begin{tabular}{c | c c c c}
                                        &  $c_{BB}$ & $c_{WW}$& $c_{WB}$&$c_{3W}$  \\ \hline
 $\gamma_{c_H}$     &  $6 g^{\prime 4}$&$18 g^{ 4}$&$6g^{\prime 2} g^2$  &      $0$      \\ 
 $\gamma_{c_T}$            &$0$&$0$&$0$           &  $0$  \\ 
$\gamma_{c_B}$           & $0$ & $0$ & $0$    &     $0$      \\ 
 $\gamma_{c_W}$       &      $0$ & $0$ & $0$   &  $2 g^2$     \\ 
 $\gamma_{c_{2B}}$      &$0$&$0$& $0$        &   $0$   \\ 
 $\gamma_{c_{2W}}$    &$0$&$0$&$0$   & $ 4 g^2 $    \\ 
 $\gamma_{c_{BB}}$     &$\frac{g^{\prime 2}}{2}-\frac{9g^2}{2}+6y_t^2+12\lambda$ & $0$ & $3 g^2$ &$0$\\ 
 $\gamma_{c_{WW}}$  &  $0$ & $-\frac{3 g^{\prime 2}}{2} - \frac{5 g^2}{2}+6y_t^2+12\lambda$ & $g^{\prime 2} $ & $ \frac{5}{2} g^2$ \\ 
 $\gamma_{c_{WB}}$   & $2 g^{\prime 2}$  & $2 g^2$ & -$\frac{g^{\prime 2}}{2}+\frac{9g^2}{2}+6y_t^2+4\lambda$ &$-\frac{g^2}{2}$ \\ 
   $\gamma_{c_{3W}}$   &$0$&$0$&$0$    & $24 g^2 - 2 \gamma_W$ \\ \hdashline
   $\gamma_{c_r}$          &   $6g^{\prime 4}$ &$18g^{ 4}$ &$6g^{\prime 2} g^2$  &   $0$         \\ 
    $\gamma_{c_{WL}^{Q,L}}$   &$0$&$0$&$0$    &  $g^2$
 \end{tabular}
\caption{\small Anomalous dimension matrix. Further contributions of ${\cal O}_{BB}, {\cal O}_{WW}, {\cal O}_{WB}$ and ${\cal O}_{3W}$ to other operators in \eq{UnivBasis1} and \eq{radred} are either zero or proportional to the Yukawa coupling of fermions lighter than the top.  The dashed line separates the anomalous dimension of the  operators in our basis from that of the redundant operators.   \label{tabLoop}}
\end{center}
\label{table:one}
\end{table}
}


{
\renewcommand{\arraystretch}{2.05} 
\renewcommand{\tabcolsep}{1.3mm}

\begin{table}[htdp]
\begin{center}
\scriptsize
\begin{tabular}{c | c c c c}
                                        &  $c_B$ & $c_W$ & $c_{2B}$ & $c_{2W}$   \\ \hline
 $\gamma_{c_H}$       &$\frac{3}{4}g^{\prime 2}\left(g^{\prime 2}+4g^2\right)$&$\frac{3}{4}g^2\left(3g^2+4g^{\prime 2}\right)-6\lambda g^2$ &$-\frac{3}{8}g^{\prime 2}\left(g^{\prime 2}+4g^2\right)$  &  $-\frac{3}{8}g^2\left(g^2 (3 + 2\xi_W) +4g^{\prime 2}\right)+3\lambda g^{2}$        \\ 
 $\gamma_{c_T}$        & $-\frac{9}{4}g^{\prime 2}g^2-6\lambda g^{\prime 2}$&$-\frac{9}{4}g^{\prime 2}g^2$& $\frac{9}{8}g^{\prime 2}g^2+3\lambda g^{\prime 2}$ & $\frac{9}{8}g^{\prime 2}g^2$ \\ 
$\gamma_{c_B}$        & $\frac{g^{\prime 2}}{6} + 6 y_t^2$&$\frac{g^2}{2}$&   $ -\frac{g^{\prime 2}}{12}$   & $-\frac{g^{ 2}}{4}$     \\ 
 $\gamma_{c_W}$       &  $\frac{g^{\prime 2}}{6}$&$\frac{11}{2}g^2+6y_t^2$&    $-\frac{g^{\prime 2}}{12}$   &   $ - g^2 \left(\frac{1}{4} + 3 \xi_W \right) $  \\  
 $\gamma_{c_{2B}}$   &$-\frac{2}{3}g^{\prime 2}$& $0$                 & $-2\gamma_{B}$ & $0$  \\ 
 $\gamma_{c_{2W}}$  &$0$&$-\frac{2}{3}g^{2}$               &  $0$& $ g^2 \left(\frac{59}{3} - 3 \xi_W \right) - 2 \gamma_{W} $ \\  
 $\gamma_{c_{BB}}$   &$0$&$0$&$0$  &$0$\\  
 $\gamma_{c_{WW}}$ &$0$&$0$ & $0$ &  $0$\\  
 $\gamma_{c_{WB}}$  &$0$&$0$& $0$ & $0$  \\  
   $\gamma_{c_{3W}}$  &$0$        &    $0$            &   $0$   & $0$ \\ \hdashline
$\gamma_{c_r}$          &$\frac{3}{2}g^{\prime 2}\left(2g^{\prime 2}-g^{ 2}\right)+6\lambda g^{\prime 2}$& $\frac{3}{2}g^{ 2}\left(6g^{ 2}-g^{\prime 2}\right)+30\lambda g^2$  & $\frac{3}{4}g^{\prime 2}\left(g^{2}-2g^{\prime 2}\right)-3\lambda g^{\prime 2}$&$-\frac{3}{4}g^{2}\left(2g^2(3 -  \xi_W) - g^{\prime 2}\right)-15\lambda g^2$          \\    
   $\gamma_{c_{K4}}$   &$-g^{\prime 2}$&$-3g^2$&    $\frac{g^{\prime 2}}{2}$   & $\frac{3}{2}g^2$           \\  
 $\gamma_{c_{L}^{(3) Q,L} }$   & $0$& $\frac{3}{4} g^4 $& $0$    &    $\frac{3}{4}g^4\xi_W$         \\ 
  $\gamma_{c_{L}^{Q,L} }$   & $0$& $0$ & $0$   &    $0$   \\ 
    $\gamma_{c_{R}^{u,d,e} }$    & $0$& $0$ & $0$ &  $0$  \\ 
 $\gamma_{c_{K3L}^{Q,L}}$  & $0$& $0$ & $- Y_F^2 g^{\prime 2}$ & $- \frac{3}{4} g^2$    \\ 
  $\gamma_{c_{K3R}^{u,d,e}}$   & $0$& $0$ &  $- Y_f^2 g^{\prime 2}$ &  $0$     \\  
   $\gamma_{c_{WL}^{Q,L}}$   & ${\cal O}\left(y_i\right)$  &  ${\cal O}\left(y_i\right)$ & $-\frac{5}{12}Y_F^2 g^{\prime 2}$ & $-\frac{21}{16} g^2 - \frac{3}{2}\xi_W g^2 $   \\ 
      $\gamma_{c_{BL}^{Q,L}}$   &${\cal O}\left(y_i\right)$ &${\cal O}\left(y_i\right)$   &   $-\frac{5}{6}Y_F^3 g^{\prime 2}$ &  $-Y_F\frac{5}{8}g^2$ \\ 
    $\gamma_{c_{BR}^{u,d,e}}$    &${\cal O}\left(y_i\right)$ &  ${\cal O}\left(y_i\right)$& $-\frac{5}{6}Y_f^3 g^{\prime 2}$  &  $0$     \\ 
    $\gamma_{c_{WL}^{\prime Q,L}}$   & ${\cal O}\left(y_i\right)$  &  ${\cal O}\left(y_i\right)$ & $-\frac{1}{2}Y_F^2 g^{\prime 2}$ &  $ - \frac{3}{8} g^2 $   \\ 
      $\gamma_{c_{BL}^{\prime  Q,L}}$  &${\cal O}\left(y_i\right)$ &${\cal O}\left(y_i\right)$   &   $-Y_F^3 g^{\prime 2}$ & $-\frac{3}{4}Y_F g^2$\\ 
    $\gamma_{c_{BR}^{\prime u,d,e}}$   &${\cal O}\left(y_i\right)$ &  ${\cal O}\left(y_i\right)$& $-Y_f^3 g^{\prime 2}$ &  $0$     \\  
           $\gamma_{c_{LL}^{(3) F}}$   & $0$ &  $0$ &  $ - \frac{3}{2} g^2 (g^{\prime }Y_F)^2 $    & $\frac{3}{8} g^2 (g^2(1+\xi_W) - 4 (g^{\prime }Y_F)^2 ) $    \\ 
       $\gamma_{c_{LL}^{F}}$   & $0$ &  $0$ &   $-6 (g^{\prime }Y_F)^4  $   &  $- \frac{9}{8} g^4 $      \\ 
      $\gamma_{c_{RR}^{f}}$   & $0$ & $0$   &  $-6 (g^{\prime }Y_f)^4 $  &  $0$ \\ 
     \end{tabular}
\caption{\small Contributions of the operators ${\cal O}_B,{\cal O}_W,{\cal O}_{2B}$ and ${\cal O}_{2W}$ to the anomalous dimension matrix of the operators in \eq{UnivBasis1} and \eq{radred}. By $y_i$ we denote the Yukawa coupling of any fermion. The dashed line separates the anomalous dimension of the operators in our basis from that of the redundant operators.   \label{tabCBW2}}
\end{center}
\end{table}
}


{
\renewcommand{\arraystretch}{2.05} 
\renewcommand{\tabcolsep}{1.3mm}

\begin{table}[htdp]
\begin{center}
\scriptsize
\begin{tabular}{c|ccccc}
 & $c_{2G}$  & $c_{GG}$ & $c_{3G}$ & $c_{2B}$ & $c_{2W}$\tabularnewline
\hline 
$ $$\gamma_{c_{2G}}$  & $\frac{1}{2}g_{s}^{2}(59-9\xi_{G})-2 \gamma_{G}$  & $0$  & $6g_{s}^{2}$  & $0$  & $0$ \tabularnewline
$\gamma_{c_{GG}}$ & $0$  & $-\frac{3}{2}g^{\prime2}-\frac{9}{2}g^{2}+12\lambda+6y_{t}^{2}$  & $0$  & $0$  & $0$ \tabularnewline
$\gamma_{c_{3G}}$ & $0$  & $0$  & $36 g_s^2 - 2 \gamma_G$ & $0$  & $0$ \tabularnewline
$\gamma_{c_{RR}^{ud}}$ & $-12 g_s^2 (g^{\prime2}Y_{u}Y_{d})$  & $0$  & $0$  & $-12(g^{\prime2}Y_{u}Y_{d})^{2}$ & $0$ \tabularnewline \hdashline
$\gamma_{c_{RR}^{(8)ud}}$ & $\frac{1}{2}g_{s}^{4}(9\xi_G-1)$  & $0$  & $0$  & $-12g_{s}^{2}(g^{\prime2}Y_{u}Y_{d})$  & $0$ \tabularnewline
$\gamma_{c_{K3L}^{Q}}$ & $-\frac{4}{3}g_{s}^{2}$  & $0$  & $0$  & $ $Table~\ref{tabCBW2}   & Table~\ref{tabCBW2} \tabularnewline
$\gamma_{c_{K3R}^{u,d}}$ & $-\frac{4}{3}g_{s}^{2}$  & $0$  & $0$  & $ $Table~\ref{tabCBW2}   & $0$ \tabularnewline
$\gamma_{c_{GR}^{u,d}}$ & $-\frac{9}{2}\xi_G-\frac{37}{9}$  & $0$  & $3 g_s^2$  & $-\frac{5}{6} (g^\prime Y_{u,d})^2$  & $0$ \tabularnewline
$\gamma_{c_{GL}^{Q}}$ & $-\frac{9}{2}\xi_G-\frac{37}{9}$  & $0$  & $3 g_s^2$  & $-\frac{5}{6} (g^\prime Y_Q)^2$  & $-\frac{5}{8} g^2$ \tabularnewline
$\gamma_{c_{WL}^{Q}}$ & $-\frac{5}{9}g_{s}^{2}$  & $0$  & $0$  & $ $Table~\ref{tabCBW2}   & Table~\ref{tabCBW2} \tabularnewline
$\gamma_{c_{BL}^{Q}}$ & $-\frac{10}{9}g_{s}^{2}Y_Q$  & $0$  & $0$  & $ $Table~\ref{tabCBW2}   & Table~\ref{tabCBW2} \tabularnewline
$\gamma_{c_{BR}^{u,d}}$ & $-\frac{10}{9}g_{s}^{2}Y_{u,d}$  & $0$  & $0$  & $ $Table~\ref{tabCBW2}   & $0$ \tabularnewline
$\gamma_{c_{GR}^{\prime u,d}}$ & $-\frac{4}{3}g_{s}^{2}$  & $0$  & $0$  & $-(g^\prime Y_{u,d})^2$  & $0$ \tabularnewline
$\gamma_{c_{GL}^{\prime Q}}$ & $-\frac{4}{3}g_{s}^{2}$  & $0$  & $0$  & $-(g^\prime Y_Q)^2$  & $-\frac{3}{4} g^2$ \tabularnewline
$\gamma_{c_{WL}^{\prime Q}}$ & $-\frac{2}{3}g_{s}^{2}$  & $0$  & $0$  & $ $Table~\ref{tabCBW2}   & Table~\ref{tabCBW2} \tabularnewline
$\gamma_{c_{BL}^{\prime Q}}$ & $-\frac{4}{3}g_{s}^{2}Y_Q$  & $0$  & $0$  & $ $Table~\ref{tabCBW2}   & Table~\ref{tabCBW2} \tabularnewline
$\gamma_{c_{BR}^{\prime u,d}}$ & $-\frac{4}{3}g_{s}^{2}Y_{u,d}$  & $0$  & $0$  & $ $Table~\ref{tabCBW2}   & $0$ \tabularnewline
\end{tabular}
\caption{\small Contributions of the operators ${\cal O}_{2G},{\cal O}_{GG},{\cal O}_{3G}, {\cal O}_{2B}$ and ${\cal O}_{2W}$ to the anomalous dimension of the operators in \eq{UnivBasis1} and \eq{radred}. 
The dashed line separates the anomalous dimension of the operators in our basis from that of the redundant operators.   \label{tabCgluons}}
\end{center}
\end{table}
}


In Table~\ref{tabCHRT} we display the contributions of ${\cal O}_H$, ${\cal O}_r$ and ${\cal O}_T$ to the running of the Wilson coefficients of the operators in \eq{UnivBasis1}. We have defined 
\be
\gamma_{c_i}= 16\pi^2 \frac{d c_i}{d \log \mu}\ ,�\quad \beta_g= \frac{dg}{d\log\mu}
\ee and 
\bea
&
\gamma_H =-N_c y_t^2+\frac{1 }{4}\left(3 [3 - \xi_W] g^2 + [3 - \xi_B]{g'}^2 \right) ,\nonumber\\[0.2cm]
&\gamma_G = - \frac{1}{g_s}\beta_{g_s} = (11 - \frac{4}{3} N_G) g_s^2 , \quad  \gamma_W = - \frac{1}{g}\beta_g =  \frac{19}{6}g^2 ,\quad  
\gamma_{B}=- \frac{1}{g^{\prime}}\beta_{g^{\prime}}=-\frac{41}{6}g^{\prime 2}  \ ,
\eea
in the background field gauge. $N_G=3$ is the number of generations. The contributions not shown are either zero or proportional to  the Yukawa coupling $y_l$ of any fermion lighter than the top. 
Notice that in Table~\ref{tabCHRT} we have gone beyond the strictly necessary computations to obtain the anomalous dimension matrix and also included the contributions of the operator ${\cal O}_r$, that is redundant with respect to our basis; their contributions are used for a crosscheck in Appendix~\ref{appenB}.

In Table~\ref{tabLoop} we show the contributions of ${\cal O}_{BB}, {\cal O}_{WW}, {\cal O}_{WB}$ and ${\cal O}_{3W}$ to the running of the operators in \eq{UnivBasis1}. The $c_{3W}$ self-renormalization has been extracted from the result of ref.~\cite{Narison:1983}. Their contribution to the running of the redundant operators in \eq{radred} that we have not written are either zero or proportional to $y_l$.

Lastly, in Table~\ref{tabCBW2}  we show the contributions of ${\cal O}_B,{\cal O}_W,{\cal O}_{2B}$ and ${\cal O}_{2W}$ to the running of any of the operators in \eq{UnivBasis1} and \eq{radred}. We have indicated by ${\cal O}\left(y_l\right)$ those contributions that \emph{at most} are expected to be proportional to the Yukawa coupling of a fermion lighter than the top. As can be noted from Table~\ref{tabCBW2}, the contribution of ${\cal O}_{2W}$ to the running of ${\cal O}_H$, ${\cal O}_r$, ${\cal O}_W$, ${\cal O}_{2W}$, $O_{L}^{(3) F_i}$, ${\cal O}_{WL}^{F_i}$ and ${\cal O}_{LL}^{(3) F_i}$ is $\xi$-dependent. This should not come as a surprise, even if we work in the background field gauge, where the counter-terms are gauge invariant. The reason is that at this point of the computation we still have redundant operators generated by the flow. By definition, in an over-complete basis that contains redundant operators only certain combinations of the Wilson coefficients enter in the physical observables. Hence, it is only after these physical combinations of the Wilson coefficients are taken, that the computation is guaranteed to be and should be gauge invariant. For instance, in Section~\ref{redred} we show that upon redefining the redundant operators in terms of operators in our basis the $\xi$ dependence of the anomalous dimension vanishes. This subtlety is well known and, for instance, it also appears in the context of Non-Relativistic QCD, where the running of the Wilson coefficients is gauge independent only when the redundancy of different operators is taken into account~\cite{Pineda:2001ra}. This has also been recently stressed again in ref.~\cite{Jenkins:2013zja}.

Table~\ref{tabCgluons} reports the contributions of ${\cal O}_{2G},{\cal O}_{GG},{\cal O}_{3G}, {\cal O}_{2B}$ and ${\cal O}_{2W}$ to the anomalous dimension of the (redundant) operators in \eq{UnivBasis1} and \eq{radred}, as needed to derive the anomalous dimension matrix of the dim-6 bosonic operators with gluons of our basis (see Table~\ref{tab:OurBasisMatrixGluons}).



 \subsection{Removal of the radiatively-generated redundant operators}
 \label{redred}
 We now turn in to discuss how to deal with each operator in \eq{radred} and their effect on the operators of \eq{UnivBasis1}.

 The easiest way to deal with the redundant operator ${\cal O}_{BR}^{\prime f_i}= g^{\prime} \widetilde{B}_{\mu \nu} i \bar{f}_{R}^i  \gamma^{\mu}  D^{\nu} f_{R}^i $~\cite{Grzadkowski:2010es} is by means of the identity\footnote{We use the conventions of \emph{Peskin \& Schroeder} textbook.}
\begin{equation}
\gamma^\mu \gamma^\nu \gamma^\rho= g^{\mu\nu} \gamma^\rho + g^{\nu\rho}\gamma^\mu - g^{\mu\rho} \gamma^\nu + i \epsilon^{\mu\nu\rho\sigma}\gamma_\sigma \gamma^5 \, ;
\label{gammasGrad}
\end{equation}
one finds 
\bea
g^{\prime} \widetilde{B}_{\mu \nu} \bar{f}_{R}  \gamma^{\mu} i D^{\nu} f_{R} &=  & \frac{g^{\prime}}{4} \bar{f}_R i \left( \gamma_\mu \gamma_\nu  \Dslash +   \overleftarrow{\Dslash}  \gamma_\mu \gamma_\nu  \right) f_R g^{\prime} \widetilde{B}^{\mu \nu}  \nonumber\\
&+  & i g^{\prime } \bar{f}_R\gamma_\rho\gamma_\mu \gamma_\nu f_R D^\rho \widetilde{B}^{\mu\nu}\, .
\label{opGrad}
\eea
Then, using the fermion's EoM
\bea
 \frac{g^{\prime}}{4}\bar{f}_R i \left( \gamma_\mu \gamma_\nu  \Dslash +   \overleftarrow{\Dslash}  \gamma_\mu \gamma_\nu  \right) f_R g^{\prime} \widetilde{B}^{\mu \nu} &=  & \frac{1}{4}g^{\prime}y_f  i \bar{F}_L \sigma_{\mu \nu}  f_R  H g^{\prime}  \widetilde{B}^{\mu\nu} + \text{h.c.}  \nonumber\\
 &=&\frac{1}{4} g^{\prime} y_f   \bar{F}_L \sigma_{\mu \nu} f_R H g^{\prime}   B^{\mu\nu} + \text{h.c.}  \equiv \frac{1}{4} {\cal O}_{DB}^f \, ,
 \label{opGrad2}
\eea
which is a \emph{dipole} operator, where $\sigma^{\mu \nu}\equiv\frac{i}{2}[\gamma^\mu,\gamma^\nu]$; using again \eq{gammasGrad} in the second term of the right hand side of \eq{opGrad}
\bea
i g^{\prime } \bar{f}_R\gamma_\rho\gamma_\mu \gamma_\nu f_R D^\rho \widetilde{B}^{\mu\nu}&= &  2 g^{\prime } \bar{f}_R\gamma_\sigma f_R D_\rho B^{\sigma\rho} = 2 {\cal O}_{BR}^f \, . 
\label{opGrad3}
\eea
Therefore, Eqs.~(\ref{opGrad})-(\ref{opGrad3}) and analogous manipulations,  are equivalent to the following shifts ($c_i\rightarrow c_i+\delta c_i$)  in the following Wilson coefficients: 
\be
\delta c_{WL}^F = 2 c_{WL}^{\prime F} \, , \quad \delta c_{BL}^F  = 2c_{BL}^{\prime F} \, , \quad  \delta c_{BR}^f = 2 c_{BR}^{\prime f} , \quad  \delta c_{GL}^Q = 2 c_{GL}^{\prime Q} , \quad  \delta c_{GR}^q = 2 c_{GR}^{\prime q}  \, .  
\label{Cshift1}
\ee
The Wilson coefficient of the dipole operators are also shifted, see \eq{opGrad2}, however, we can not conclude that the dipoles are renormalized by the set of bosonic operators we considered because we did not compute direct contributions, those coming from one-particle-irreducible diagrams.

Then, for the operator ${\cal O}_{K3R}^{f_i}$, consider the field redefinition $\delta f_i= -\frac{c_{K3R}^{f_i}}{2\Lambda^2} D^2 f_i $, that removes ${\cal O}_{K3R}^{f_i}$ from the Lagrangian while generates the operator 
\begin{equation}
\begin{split}
- \frac{c_{K3R}^{f_i} y_{f_i}}{2\Lambda^2} D_\mu \bar{F}_{iL} D^{\mu} \left( f_{i R} H \right)+\text{h.c.} = - \frac{c_{K3R}^{f_i} y_{f_i}}{2\Lambda^2}&[  D_\mu \bar{F}_{iL}\gamma^\mu \gamma^\nu D^{\nu} \left( f_{i R} H \right)\\
& -\frac{1}{2} \bar{F}_{iL}X_{\mu \nu} \sigma^{\mu \nu}  f_{i R} H  +\text{h.c.} ]\, ,
\label{mbeitw}
\end{split}
\end{equation}
where $X_{\mu \nu}=g^{\prime} Y_{F_i} B_{\mu\nu}+ gW^a_{\mu \nu}\tau^a + g_s G^A_{\mu\nu}T^a$, being $\tau^a$ and $T^A$ the $SU(2)_L$ and $SU(3)_c$ generators in the fundamental representation, respectively. Then, by inserting the fermion's EoM in the first operator in the right hand side of \eq{mbeitw} one gets operators of the type ${\cal L}_{\text{Yuk}}\left|H\right|^2$ and the operator $y_{f_i}{\cal O}_R^{f_i}\equiv y_{f_i}i( H^\dagger {\lra { D_\mu}} H)  \bar{f}_{R}^i \gamma^{\mu} f_{R}^i$; we do not care about the latter (proportional to $y_{f_i}$) since our basis choice of Section~\ref{Dim6Basis} was to remove the operator ${\cal O}_{R}^{f_i}$ corresponding to a light fermion. Performing an analogous analysis for ${\cal O}_{K3L}^{F_i}$ we reach the same conclusion: neither of the two operator's scaling affects the anomalous dimension of the set of bosonic operators in \eq{UnivBasis1}.  As in the case of ${\cal O}^{\prime }_{WL,BL,BR}$, the same comment applies here: even-though the Wilson coefficient of the dipoles is shifted by the above manipulations, we do not conclude that they are renormalized by the bosonic operators.


Now, the remaining operators (corresponding to the third, forth and fifth line of \eq{radred}) are redefined into our basis by performing field redefinitions. 
Consider the 37 independent field redefinitions
\be \begin{split}
	\Lambda^2 \delta G_\mu^A   & =  \alpha_{2G}  (D^\nu G^A_{\mu\nu}) + g_S \sum_i \alpha_{QG}^i \bar{Q^i}_L T^A \gamma_\mu Q^i_L + g_S \sum_{i, q} \alpha_{qG}^i \bar{q^i}_R T^A \gamma_\mu q^i_R,, \\
	\Lambda^2 \delta W_\mu^a   & =  i g \alpha_W  (H^\dagger\sigma^a \lra{D^\mu} H) +  \alpha_{2W}  (D^\nu W^a_{\mu\nu}) + g \sum_{i, F} \alpha_{FW}^i \bar{F^i}_L \sigma^a \gamma_\mu F^i_L, \\
	\Lambda^2 \delta B_\mu  & = i g^{\prime } \alpha_B  (H^\dagger \lra{D^\mu} H) +  \alpha_{2B}  (\partial^\nu B_{\mu\nu})  + g^{\prime } \sum_{i,F} Y_F \alpha_{FB}^i \bar{F}_L^i \gamma_\mu F_L^i + g^{\prime} \sum_{i,f} Y_f \alpha_{fB}^i \bar{f}_R^i \gamma_\mu f_R^i, \\
	\Lambda^2 \delta H   & = \alpha_1 H |H|^2+\alpha_2 \left( (D^2 H)  - y_{e}^{ij}  \bar{e}_R^i L_L^j  - y_{d}^{ij}  \bar{d}_R^i Q_L^j - y_{u}^{ij} i \sigma^2 (\bar{u}_R^i Q_L^j)^* \right),
	\label{eq:AllFieldRed}
\end{split} \ee
with $F= \{L,Q\}$, $f = \{ e,d,u\}$, $q = \{d,u\}$ and $i=1,2,3$.
These generate the following shifts for the Wilson coefficients of the dimension 6 operators: 
\be
\begin{array}{rl rl}
c_{H}\rightarrow& c_{H}+2(\alpha_{1}+2\lambda\alpha_{2})-\alpha_{W}g^{2} & c_{r}\rightarrow& c_{r}+2(\alpha_{1}+2\lambda\alpha_{2})+\alpha_{W}g^{2} \vspace{0.5em} \\
c_{T}\rightarrow& c_{T}-\alpha_{B}g^{\prime2} & c_{K4}\rightarrow& c_{K4}-2\alpha_{2} \vspace{0.5em} \\
c_{B}\rightarrow& c_{B}+\alpha_{2B}-2\alpha_{B} & c_{WL}^{F_{i}}\rightarrow& c_{WL}^{F_{i}}+\frac{1}{2}\alpha_{2W}-\alpha_{FW}^{i} \vspace{0.5em} \\
c_{W}\rightarrow& c_{W}+\alpha_{2W}-2\alpha_{W} & c_{BL}^{F_{i}}\rightarrow& c_{BL}^{F_{i}}+Y_{F}(\alpha_{2B}-\alpha_{FB}^{i}) \vspace{0.5em} \\
c_{2B}\rightarrow& c_{2B}+2\alpha_{2B} & c_{BR}^{f{}_{i}}\rightarrow& c_{BR}^{f{}_{i}}+Y_{f}(\alpha_{2B}-\alpha_{fB}^{i}) \vspace{0.5em} \\
c_{2W}\rightarrow& c_{2W}+2\alpha_{2W} & c_{LL}^{(3)F_{i}}\rightarrow& c_{LL}^{(3)F_{i}}+\frac{g^{2}}{2}\alpha_{FW}^{i} \vspace{0.5em} \\
c_{2G}\rightarrow& c_{2G} + 2\alpha_{2G} & c_{LL}^{F_{i}}\rightarrow& c_{LL}^{F_{i}}+(Y_{F}g^{\prime})^{2}\alpha_{FB}^{i} \vspace{0.5em} \\
c_{6}\rightarrow& c_{6}-4\alpha_{1} & c_{RR}^{f{}_{i}}\rightarrow& c_{RR}^{f{}_{i}}+(Y_{f}g^{\prime})^{2}\alpha_{fB}^{i} \vspace{0.5em} \\
c_{y_{f}}^{i}\rightarrow& c_{y_{f}}^{i}-\alpha_{1}+2\lambda\alpha_{2} & c_{LR}^{F_{i}f{}_{j}}\rightarrow& c_{LR}^{F_{i}f{}_{j}}+(Y_{F}Y_{f}g^{\prime2})(\alpha_{fB}^{i}+\alpha_{FB}^{i}) \vspace{0.5em} \\
c_{y_{f}y_{f}}^{ij}\rightarrow& c_{y_{f}y_{f}}^{ij}+2\alpha_{2}  & c_{L}^{(3)F_{i}}\rightarrow& c_{L}^{(3)F_{i}}+\frac{g^{2}}{2}(\alpha_{W}+\alpha_{FW}^{i}) \vspace{0.5em} \\
 c_{RR}^{u^i d^j} \rightarrow& c_{RR}^{u^i d^j} + g^{\prime 2}Y_u Y_d (\alpha_{uB}^i + \alpha_{dB}^j) & c_{L}^{F_{i}}\rightarrow& c_{L}^{F_{i}}+Y_{F}g^{\prime2}(\alpha_{B}+\frac{1}{2}\alpha_{FB}^{i}) \vspace{0.5em} \\
& & c_{R}^{f{}_{i}}\rightarrow& c_{R}^{f{}_{i}}+Y_{f}g^{\prime2}(\alpha_{B}+\frac{1}{2}\alpha_{fB}^{i}) \\
& & c_{GL,R}^{q_i} \rightarrow& c_{GL,R}^{q_i} + \alpha_{2G} - \alpha_{qG}^i \; \text{for }q=Q,u,d \\
& & c_{RR}^{(8)u^i d^j} \rightarrow& c_{RR}^{(8)u^i d^j} + g_s^2 ( \alpha_{uG}^i + \alpha_{dG}^j ).
\end{array}
\label{eq:AllShifts}
\ee
Notice that using Fierz identities we can always trade the operator $\Op_{LL}^{F_i}$ for $\Op_{LL}^{(3) F_i}$: $\Op_{LL}^{F_i} = \Op_{LL}^{(3) F_i}$. This means that the shift in $c_{LL}^{F_i}$ can be recast as a shift in $c_{LL}^{(3) F_i}$, which becomes:
\be
	c_{LL}^{(3)F_{i}}\rightarrow c_{LL}^{(3)F_{i}}+\frac{g^{2}}{2}\alpha_{FW}^{i} + \left( c_{LL}^{F_{i}}+(Y_{F}g^{\prime})^{2}\alpha_{FB}^{i} \right).
\ee
 We use the freedom given by the field redefinitions to set to zero the following 37 coefficients: $c_r, c_{K4}, c_{LL}^{(3) L_1}, c_{RR}^{e_1}, c_{L}^{(3) L_1}, c_{R}^{e_1}, c_{WL}^{F_i}, c_{BL}^{F_i}, c_{BR}^{f_i}, c_{GL}^{Q_i}, c_{GR}^{u_i}, c_{GR}^{d_i}, c_{RR}^{(8)u^1 d^1}$. This fixes all the shift parameters $\alpha_i$ and gives shift invariant combinations, under  \eq{eq:AllShifts}, of the Wilson coefficients of the operators in our basis:
\be\begin{split}
c_{H} & \rightarrow  c_{H}-c_{r}+6(c_{L}^{(3)L_{1}}-\tilde{c}_{LL}^{(3)L_{1}})\ ,\\
c_{T} & \rightarrow  c_{T}+\frac{1}{Y_{e}}(c_{R}^{e_{1}}-\frac{1}{2Y_{e}}c_{RR}^{e_{1}})\ ,\\
c_{W} & \rightarrow c_{W}-2c_{WL}^{L_{1}}-4c_{WL}^{\prime L_{1}}+\frac{4}{g^{2}}(c_{L}^{(3)L_{1}}-2\tilde{c}_{LL}^{(3)L_{1}})\ ,\\
c_{B} & \rightarrow  c_{B}-\frac{1}{Y_{e}}c_{BR}^{e_{1}}-\frac{2}{Y_{e}}c_{BR}^{\prime e_{1}}+\frac{2}{Y_{e}g^{\prime2}}(c_{R}^{e_{1}}-\frac{1}{Y_{e}}c_{RR}^{e_{1}})\ ,\\
c_{2W} & \rightarrow  c_{2W}-4c_{WL}^{L_{1}}-8c_{WL}^{\prime L_{1}}-\frac{8}{g^{2}}\tilde{c}_{LL}^{(3)L_{1}}\ ,\\
c_{2B} & \rightarrow  c_{2B}-\frac{2}{Y_{e}}c_{BR}^{e_{1}}-\frac{4}{Y_{e}}c_{BR}^{\prime e_{1}}-\frac{2}{Y_{e}^{2}g^{\prime2}}c_{RR}^{e_{1}} \, ,\\
c_{6} & \rightarrow  c_{6}+2c_{r}+4\lambda c_{K4}-8(c_{L}^{(3)L_{1}}-\tilde{c}_{LL}^{(3)L_{1}}) \; , \\
c_{2G} & \rightarrow  c_{2G} - c_{GR}^{d_1}  - 2 c_{GR}^{\prime d_1} - c_{GR}^{u_1} - 2 c_{GR}^{\prime u_1} - \frac{1}{g_s^2} c_{RR}^{(8) u^1 d^1} \ ,
\label{eq:PhysicalWilsComb}
\end{split}\ee
where
\be
	\tilde{c}_{LL}^{(3)L_{1}} = c_{LL}^{(3)L_{1}} + c_{LL}^{F_1} + g^{\prime 2} Y_L \left(  c_{BL}^{L_1} + 2 c_{BL}^{\prime L_1} - \frac{Y_L}{Y_e} (c_{BR}^{e_1} + 2 c_{BR}^{\prime e_1} + \frac{1}{g^{\prime 2} Y_e} c_{RR}^{e_1} ) \right) \, 
\ee
and \eq{Cshift1} has already been taken into account. This completes the removal of the operators in \eq{radred} in terms of the bosonic operators.

As we have just shown,  upon eliminating the redundant operators the Wilson coefficients of the operators of \eq{UnivBasis1} are shifted in such a way that the anomalous dimensions are redefined as
\begin{equation} \begin{split}
	\gamma_{c_H} & \rightarrow  \gamma_{c_H}-\gamma_{c_r}+6(\gamma_{c_{L}^{(3)L_{1}}}-\tilde{\gamma}_{c_{LL}^{(3)L_{1}}}) \ ,\\
	\gamma_{c_T} & \rightarrow  \gamma_{c_T}+\frac{1}{Y_{e}}(\gamma_{c_{R}^{e_{1}}}-\frac{1}{2Y_{e}}\gamma_{c_{RR}^{e_{1}}})\ ,\\
	\gamma_{c_W} & \rightarrow  \gamma_{c_W}-2\gamma_{c_{WL}^{L_{1}}}-4\gamma_{c_{WL}^{\prime L_{1}}}+\frac{4}{g^{2}}(\gamma_{c_{L}^{(3)L_{1}}}-2\tilde{\gamma}_{c_{LL}^{(3)L_{1}}})\ ,\\
	\gamma_{c_B} & \rightarrow  \gamma_{c_B}-\frac{1}{Y_{e}}\gamma_{c_{BR}^{e_{1}}}-\frac{2}{Y_{e}}\gamma_{c_{BR}^{\prime e_{1}}}+\frac{2}{Y_{e}g^{\prime2}}(\gamma_{R}^{e_{1}}-\frac{1}{Y_{e}}\gamma_{c_{RR}^{e_{1}}})\ ,\\
	\gamma_{c_{2W}} & \rightarrow  \gamma_{c_{2W}}-4\gamma_{c_{WL}^{L_{1}}}-8\gamma_{c_{WL}^{\prime L_{1}}}-\frac{8}{g^{2}}\tilde{\gamma}_{c_{LL}^{(3)L_{1}}} \ ,\\
	\gamma_{c_{2B}} & \rightarrow  \gamma_{c_{2B}}-\frac{2}{Y_{e}}\gamma_{c_{BR}^{e_{1}}}-\frac{4}{Y_{e}}\gamma_{c_{BR}^{\prime e_{1}}}-\frac{2}{Y_{e}^{2}g^{\prime2}}\gamma_{c_{RR}^{e_{1}}}\  ,\\
	\gamma_{c_6} & \rightarrow  \gamma_{c_6}+2\gamma_{c_{r}}+4\lambda \gamma_{c_{K4}}-8(\gamma_{c_{L}^{(3)L_{1}}}-\tilde{\gamma}_{c_{LL}^{(3)L_{1}}})\  , \\
	\gamma_{c_{2G}} & \rightarrow \gamma_{c_{2G}} - \gamma_{c_{GR}^{d_1}} - \gamma_{c_{GR}^{u_1}} - \frac{1}{g_s^2} \gamma_{c_{RR}^{(8) u^1 d^1}}   \ ,
\label{physGamma}
\end{split}\end{equation}
where
\be
	\tilde{\gamma}_{c_{LL}^{(3)L_{1}}} = \gamma_{c_{LL}^{(3)L_{1}}} + \gamma_{c_{LL}^{F_1}} + g^{\prime 2} Y_L \left(  \gamma_{c_{BL}^{L_1}} + 2 \gamma_{c_{BL}^{\prime L_1}} - \frac{Y_L}{Y_e} (\gamma_{{c_BR}^{e_1}} + 2 \gamma_{c_{BR}^{\prime e_1}} + \frac{1}{g^{\prime 2} Y_e} \gamma_{c_{RR}^{e_1}} ) \right).
\ee
after setting $c_{red}^{0, i}=0$, being $c_{red}^{0,i}$ the tree-level Wilson coefficient of any of the operators in \eq{radred}. The anomalous dimensions of the remaining bosonic operators, that are not of the form \emph{(SM current)}$\times$\emph{(SM current)}, are not redefined.
In this way we can go back to our original basis taking into account that some operators are generated radiatively even if we set their Wilson coefficient to zero at the matching scale. In the main body of the paper, Tables~\ref{tab:OurBasisMatrix} and~\ref{tab:OurBasisMatrixGluons}, we gave the physical anomalous dimensions obtained using the right hand side of \eq{physGamma}. As  announced in Section~\ref{appenA1}, the $\xi$ dependence cancels out in the physical combinations of $\gamma_{c_i}$'s, which can be easily checked using \eq{physGamma}.

\section{Field Reparametrization-Invariance Crosscheck}
\label{appenB}

There is a useful consistency check that can be done to the results presented in Tables~\ref{tabCHRT} and~\ref{tabCBW2}. Consider the set of 9~operators
\be
{\cal B}=\left\{{\cal O}_{K4}, {\cal O}_6, {\cal O}_H, {\cal O}_r, {\cal O}_T, {\cal O}_B, {\cal O}_W, {\cal O}_{2B}, {\cal O}_{2W} \right\} \ .
\label{eq:NoFermOperators}
\ee
By means of field redefinitions, these operators are related among themselves and to other operators that contain fermions, see \eq{eq:AllFieldRed}. Therefore, in a hypothetical theory with no fermions, but otherwise equivalent to the SM, the operator set of \eq{eq:NoFermOperators} would be over-complete, i.e. there would be operators which could be removed using field redefinitions. Let us take this scenario as a working assumption for the rest of this Appendix.
More concretely, consider the subset of field redefinitions of \eq{eq:AllFieldRed}, parametrized by 
\be
\{ \alpha_1, \, \alpha_2, \, \alpha_B, \, \alpha_{2B}, \, \alpha_W, \, \alpha_{2W} \}
\label{eq:ShiftParNoFerm}
\ee
and the shifts they produce on the operators of \eq{eq:NoFermOperators} given in \eq{eq:AllShifts}.
Using this shift freedom we can choose to remove all the operators in $\cal B$ except ${\cal O}_6$, ${\cal O}_H$ and ${\cal O}_T$. 
However, notice that the over-completeness\footnote{Again,  we stress that the set of operators in \eq{eq:NoFermOperators} is over-complete  only in the absence of the SM fermions. } of ${\cal B}$ can be exploited in our advantage; physical observables are independent of the coordinates choice as long as such a choice is compatible with the assumed symmetries. {Hence, physical observables can not depend on the arbitrary parameters $\alpha_i$ of \eq{eq:ShiftParNoFerm} that we used to parametrize the field redefinitions. The following combinations of Wilson coefficients are invariant under such shifts:
\bea
C_H&\equiv & c_H - c_r -\frac{3}{4}g^2(2c_W-c_{2W})\ ,\nonumber\\
C_T&\equiv & c_T-\frac{1}{4}{g'}^2(2c_B-c_{2B})\ , \\
C_6&\equiv & c_6 +2c_r+g^2(2c_W-c_{2W})+4\lambda c_{K4} \nonumber \ .
\label{physcoef}
\eea
Physical observables depend on shift invariant combinations of couplings, which we denote by a capital \emph{$C_i$}. Also, a key property is that the anomalous dimension of a shift invariant combination of couplings is a function of shift invariant combinations of couplings only \footnote{See ref.~\cite{Elias-Miro:2013gya} for a more detailed discussion. }
\be
\gamma_{C_i} = f(C_j)\ .
\label{gamma}
\ee
This is precisely the cross-check that can be done to the results computed in Tables~\ref{tabCHRT} and~\ref{tabCBW2}. And indeed it is easy to check that:
\be \begin{split}
\gamma_{C_H} &=\; \left( 24 \lambda - 4 g^2 -  3 g^{\prime 2} \right)  C_H + \frac{1}{2}\left(  24 \lambda +  9 g^{\prime 2} -17g^2\right) C_T , \\
\gamma_{C_T} &=\;  \frac{1}{6} \left( 72\lambda +5 g^{\prime 2} +27g^2 \right) C_T + \frac{5}{3}g^{\prime 2}C_H \ ,
\label{physGamma2}
\end{split} \ee
 as it should, given the fact that ${\cal O}_6$ does not renormalize ${\cal O}_r$, ${\cal O}_{H}$, ${\cal O}_{T}$. As its clear from the discussion above, to compute \eq{physGamma2} one has to insert the Higgs and gauge bosons anomalous dimensions and the gauge beta functions without the contributions of the fermions:
\be
\gamma_H^{nf} =\gamma_H |_{y_f=0} \ ,\quad \gamma_W^{nf} = - \frac{1}{g}\beta_g^{nf} =  \frac{43}{6}g^2 \ ,\quad  
\gamma_{B}^{nf}=-\frac{1}{g^{\prime}} \beta_{g^{\prime}}^{nf}=-\frac{g^{\prime 2}}{6}  \ ,
\ee
in the background field gauge and the superscript \emph{nf} stands for \emph{no fermions}, to distinguish them from their SM counterparts. 

Notice also that in \eq{physGamma2} the $\xi$ dependence exactly cancels, as it should, rendering the result independent of the gauge fixing term of \eq{bkgFG}.

\section{Comparison with previous literature}
\label{appenHagiwara}

Let us now put into context the results for the anomalous dimensions presented in this paper. The first paper in the literature with a similar spirit to ours is ref.~\cite{Hagiwara:1993ck}, followed by ref.~\cite{Alam:1997nk} and more recently by refs.~\cite{Mebane:2013zga} and~\cite{Chen:2013kfa}, where they present the contributions of the operators~\footnote{Notice that we have chosen different normalizations for the operators, different conventions for the covariant derivatives and different names for the operators with respect to ref.~\cite{Hagiwara:1993ck}.} 
\be
 {\cal O}_{3W}  ,  \  {\cal O}_{WW} , \ {\cal O}_{BB}  ,  \ {\cal O}_{HW}\equiv ig (D^\mu H)^\dagger \sigma^a (D^\nu H) W^a_{\mu\nu}  ,  \ {\cal O}_{HB}\equiv ig^\prime(D^\mu H)^\dagger  (D^\nu H) B_{\mu\nu} \ ,
 \label{HagiwaraOps}
\ee
to the running of the Wilson coefficients of the operators $\{{\cal O}_{2W} ,  {\cal O}_{2B} , {\cal O}_{WB} , {\cal O}_{T}\}$ \footnote{In fact they do not consider ${\cal O}_{T}$ but ${\cal O}_{\Phi,1}\equiv({\cal O}_H-{\cal O}_T)/2$. However only the projection of ${\cal O}_{\Phi,1}$ into ${\cal O}_T$ enters in the T-parameter.}. The results in formulas $(4.9\text{a})-(4.9\text{d})$ of ref.~\cite{Hagiwara:1993ck} recast in terms of our conventions are shown in Table~\ref{tab:OurBasisMatrix2}; they correspond to the contributions of $\{ c_{HB},c_{HW}, c_{BB}, c_{3W}, c_{WW}\}$ to $\{\gamma_{c_{2B}}, \gamma_{c_{2W}},\gamma_{c_{WB}},\gamma_{c_T}\}$, we find complete agreement.

We want to stress here that the approach we followed to compute the running of the Wilson coefficients is somewhat different than in ref.~\cite{Hagiwara:1993ck}. We computed the effective action (only one-particle irreducible graphs) in the background field gauge starting from a basis of operators; then, along the RG flow operators not included in the basis (like ${\cal O}_L^{(3)}$) are generated. These are redefined into our basis, and they are interpreted as the vertex ($e_L\sigma^a \gamma^\mu e_L W^a_\mu$) corrections that are computed in refs.~\cite{Hagiwara:1994pw,Hagiwara:1993ck} to make the result physical and hence gauge invariant. We believe that the approach we followed to compute the anomalous dimensions is somewhat more systematic when dealing with the renormalization of other operators than the oblique ones. 

\medskip

Contributions from ${\cal O}_H$ to the T and S-parameters  are given in ref.~\cite{Barbieri:2007bh}. Then, the separate contributions of $\{{\cal O}_T, {\cal O}_H\}$ to $\{{\cal O}_B, {\cal O}_W\}$ are given in ref.~\cite{Elias-Miro:2013gya}.

The $3\times 3$ matrix of anomalous dimensions for the operators $\{{\cal O}_{BB}, {\cal O}_{WB}, {\cal O}_{WW}\}$ has been computed in ref.~\cite{Grojean:2013kd}, together with its CP-odd counterparts. 
Then, in ref.~\cite{Elias-Miro:2013gya} it is shown that the $3\times3$ anomalous dimensions matrix computed in ref.~\cite{Grojean:2013kd} does not mix with the $2 \times 2$ anomalous dimension matrix of the operators $\{ {\cal O}_B, {\cal O}_W \}$. This later result, together with the use of the EOM or field redefinitions implies that none of the $\text{\emph{(SM current)}}\times\text{\emph{(SM current)}}$ dim-6 operators renormalizes the operators $\{{\cal O}_{BB}, {\cal O}_{WB}, {\cal O}_{WW}\}$. 
In ref.~\cite{Elias-Miro:2013gya} the contribution of dipole operators (like ${\cal O}_{DB}$ in Appendix~\ref{redred}) to the operators $\{{\cal O}_{BB}, {\cal O}_{WB}, {\cal O}_{WW}\}$, and to its CP-odd counterparts, is also computed. 

In ref.~\cite{EEMP2} several anomalous dimensions were computed, some of them overlap with the work presented here. These are the contributions of $\{ {\cal O}_H, {\cal O}_r\} $ to the anomalous dimension of any dim-6 operator. The contributions from operators containing fermions to the anomalous dimensions of any interesting operator for Higgs physics or EW precision tests are also computed in ref.~\cite{EEMP2}.  

The Yukawa dependence of the anomalous dimensions matrix of the dim-6 SM operators is given in ref.~\cite{Jenkins:2013wua}. However, notice the Yukawa dependences needed in the present paper to derive RG-induced constraints come only from the wave functions of the SM particle's field or are proportional to small Yukawas. 

\bigskip

Tables~\ref{tab:OurBasisMatrix2} and~\ref{tab:OurBasisMatrix3}  show the results for the anomalous dimensions matrix presented in the main body of the paper in two of the most used bases in the literature, refs.~\cite{Hagiwara:1993ck} and~\cite{Giudice:2007fh}. The three bases differ in the choice of 5 bosonic operators among the redundant set of the 7~operators $\{ {\cal O}_{BB} ,  {\cal O}_{WW}  ,   {\cal O}_{WB} ,  {\cal O}_{HB},  {\cal O}_{HW},  {\cal O}_{B},  {\cal O}_{W} \}$:
 ref.~\cite{Hagiwara:1993ck} drops the 2 operators $\{ {\cal O}_{B},  {\cal O}_{W} \}$ while ref.~\cite{Giudice:2007fh} does not use the 2 operators 
 $\{ {\cal O}_{WW}  ,   {\cal O}_{WB} \}$, and our basis leaves out the 2 operators $\{  {\cal O}_{HB},  {\cal O}_{HW} \}$.
The three bases are connected by means of the identities
\bea
&& {\cal O}_W={\cal O}_{HW} +\frac{1}{4}\left({\cal O}_{WW}+{\cal O}_{WB}\right)\ , \nonumber\\
&& {\cal O}_B={\cal O}_{HB}+\frac{1}{4}\left({\cal O}_{WB}+{\cal O}_{BB}\right)\ . 
\label{treeLoopID}
\eea
A good property of our basis with respect to the ones in the literature is that the one-loop anomalous dimension matrix is simpler, since its has a block diagonal structure.  

{
\renewcommand{\arraystretch}{1.55} 
\renewcommand{\tabcolsep}{2mm}

\begin{table}[htdp]
\begin{center}
\scriptsize

\begin{tabular}{C{0.8cm}|c c}
 & $c_{H}$  & $c_{T}$ \\
\hline 
$\gamma_{c_{H}}$  & $-\frac{9}{2}g^{2}-3g^{\prime2}+24\lambda+12y_{t}^{2}$  & $-9g^{2}+\frac{9}{2}g^{\prime2}+12\lambda$ \\
$\gamma_{c_{T}}$  & $\frac{3}{2}g^{\prime2}$  & $\frac{9}{2}g^{2}+12\lambda+12y_{t}^{2}$ \\
$\gamma_{c_{2B}}$  & $0$ & $0$\\
$\gamma_{c_{2W}}$  & $0$ & $0$\\
$\gamma_{c_{HB}}$  & $-\frac{1}{3}$  & $-\frac{5}{3}$ \\
$\gamma_{c_{HW}}$  & $-\frac{1}{3}$  & $-\frac{1}{3}$ \\
$\gamma_{c_{BB}}$  & $-\frac{1}{12}$ & $-\frac{5}{12}$ \\
$\gamma_{c_{WW}}$  & $-\frac{1}{12}$ &  $-\frac{1}{12}$ \\
$\gamma_{c_{WB}}$  & $-\frac{1}{6}$ & $-\frac{1}{2}$ \\
$\gamma_{c_{3W}}$  & $0$ & $0$ 
\end{tabular}

\vspace{0.5cm}

\begin{tabular}{C{0.7cm} | C{3.115cm} C{3.115cm} C{3.115cm} C{3.115cm} }
 & $c_{2B}$  & $c_{2W}$  & $c_{HB}$  & $c_{HW}$ \tabularnewline
\hline 
$\gamma_{c_{H}}$  & $-\frac{141}{16}g^{\prime4}+3g^{\prime2}\lambda$  & $\frac{63}{8}g^{4}-\frac{51}{16}g^{2}g^{\prime2}+18\lambda g^{2}$  & $-\frac{9}{4}g^{\prime2}(g^{\prime2}-2g^{2})-6\lambda g^{\prime2}$ & $-\frac{9}{4}g^{2}(g^{2}-2g^{\prime2})-36\lambda g^{\prime2}$\tabularnewline
$\gamma_{c_{T}}$  & $3g^{\prime4}+\frac{9}{8}g^{\prime2}g^{2}+3\lambda g^{\prime2}$  & $\frac{9}{8}g^{\prime2}g^{2}$  & $-\frac{9}{4}g^{\prime2}g^{2}-6\lambda g^{\prime2}$ & $-\frac{9}{4}g^{\prime2}g^{2}$\tabularnewline
$\gamma_{c_{2B}}$  & $\frac{94}{3}g^{\prime2}$ & $0$  & $-\frac{2}{3}g^{\prime2}$ & $0$ \tabularnewline
$\gamma_{c_{2W}}$  & $\frac{53}{12}g^{\prime2}\left(1-3t_{\theta_{W}}^{2}\right)$ & $\frac{331}{12}g^{2}+\frac{29}{4}g^{\prime2}$  & $0$ & $-\frac{2}{3}g^{2}$ \tabularnewline
$\gamma_{c_{\text{\emph{HB}}}}$  & \textbf{$\frac{59}{4}g^{\prime2}$} & $-\frac{g^{2}}{4}$  & $\frac{g^{\prime2}}{6}+6y_{t}^{2}$ & $\frac{g^{2}}{2}$\tabularnewline
$\gamma_{c_{\text{\emph{HW}}}}$  & $\left(\frac{29}{8}-\frac{53}{4}t_{\theta_{W}}^{2}\right)g^{\prime2}$  & $\frac{79}{8}g^{2}+\frac{29}{4}g^{\prime2}$  & $\frac{g^{\prime2}}{6}$ & $\frac{17}{2}g^{2}+6y_{t}^{2}$\tabularnewline
$\gamma_{c_{BB}}$  & \textbf{$\frac{59}{16}g^{\prime2}$} & $-\frac{1}{16}g^{2}$ & $\frac{3}{8}g^{2}-\frac{1}{12}g^{\prime2}-3\lambda$ & $-\frac{5}{8}g^{2}$\tabularnewline
$\gamma_{c_{\text{\emph{WW}}}}$  & $\frac{1}{4}\left(\frac{29}{8}-\frac{53}{4}t_{\theta_{W}}^{2}\right)g^{\prime2}$  & $\frac{79}{32}g^{2}+\frac{29}{16}g^{\prime2}$  & $-\frac{5}{24}g^{\prime2}$ & $\frac{11}{4}g^{2}+\frac{1}{8}g^{\prime2}-3\lambda$ \tabularnewline
$\gamma_{c_{\text{\emph{WB}}}}$  & $\frac{1}{4}\left( \frac{147}{8}-\frac{53}{4}t_{\theta_{W}}^{2}\right)g^{\prime2}$ & $ \frac{77}{32}g^{2}+\frac{29}{16}g^{\prime2}$ & $-\frac{9}{8}g^{2}-\frac{7}{24}g^{\prime2}-\lambda$ & $\frac{5}{8}g^{2}+\frac{1}{8}g^{\prime2}-\lambda$\tabularnewline
$\gamma_{c_{3W}}$  & $0$  & $0$  & $0$  & $0$ \tabularnewline
\end{tabular}

\vspace{0.5cm}

\begin{tabular}{C{0.8cm} | C{3.115cm} C{3.115cm} C{3.115cm} C{3.115cm} }
 & $c_{BB}$  & $c_{WW}$ & $c_{WB}$ & $c_{3W}$ \tabularnewline
\hline 
others  & $0$  & $0$  & $0$  & $0$ \tabularnewline
$\gamma_{c_{BB}}$  & $\frac{g^{\prime2}}{2}-\frac{9g^{2}}{2}+6y_{t}^{2}+12\lambda$  & $0$  & $3g^{2}$  & $0$\tabularnewline
$\gamma_{c_{\text{\emph{WW}}}}$  & $0$  & $-\frac{3g^{\prime2}}{2}-\frac{5g^{2}}{2}+6y_{t}^{2}+12\lambda$  & $g^{\prime2}$  & $\frac{5}{2}g^{2}$ \tabularnewline
$\gamma_{c_{WB}}$  & $2g^{\prime2}$  & $2g^{2}$  & $\frac{9g^{2}}{2}-\frac{g^{\prime2}}{2}+6y_{t}^{2}+4\lambda$  & $-\frac{g^{2}}{2}$ \tabularnewline
$\gamma_{c_{3W}}$  & $0$ & $0$ & $0$  & $\frac{53}{3} g^2$ \tabularnewline
\end{tabular}
\caption{\small Anomalous dimension matrix for the Wilson coefficients of the dim-6 bosonic operators, in the Hagiwara et. al. basis~\cite{Hagiwara:1993ck}. }
\label{tab:OurBasisMatrix2}
\end{center}
\end{table}
}

{
\renewcommand{\arraystretch}{1.55} 
\renewcommand{\tabcolsep}{2mm}

\begin{table}[htdp]
\begin{center}
\scriptsize

\begin{tabular}{C{0.8cm}|c c}
 & $c_{H}$  & $c_{T}$ \\
\hline 
$\gamma_{c_{H}}$  & $-\frac{9}{2}g^{2}-3g^{\prime2}+24\lambda+12y_{t}^{2}$  & $-9g^{2}+\frac{9}{2}g^{\prime2}+12\lambda$ \\
$\gamma_{c_{T}}$  & $\frac{3}{2}g^{\prime2}$  & $\frac{9}{2}g^{2}+12\lambda+12y_{t}^{2}$ \\
$\gamma_{c_{2B}}$  & $0$ & $0$\\
$\gamma_{c_{2W}}$  & $0$ & $0$\\
$\gamma_{c_{B}}$  & $-\frac{1}{3}$  & $-\frac{5}{3}$ \\
$\gamma_{c_{W}}$  & $-\frac{1}{3}$  & $-\frac{1}{3}$ \\
$\gamma_{c_{BB}}$  & $0$ & $0$ \\
$\gamma_{c_{HB}}$  & $0$ &  $0$ \\
$\gamma_{c_{HW}}$  & $0$ & $0$ \\
$\gamma_{c_{3W}}$  & $0$ & $0$ 
\end{tabular}

\vspace{0.5cm}

\begin{tabular}{C{0.8cm} | C{3.115cm} C{3.115cm} C{3.115cm} C{3.115cm} }
 & $c_{2B}$  & $c_{2W}$  & $c_{B}$  & $c_{W}$ \tabularnewline
\hline 
$\gamma_{c_{H}}$  & $-\frac{141}{16}g^{\prime4}+3g^{\prime2}\lambda$  & $\frac{63}{8}g^{4}-\frac{51}{16}g^{2}g^{\prime2}+18\lambda g^{2}$  & $-\frac{9}{4}g^{\prime2}(g^{\prime2}-2g^{2})-6\lambda g^{\prime2}$ & $-\frac{9}{4}g^{2}(g^{2}-2g^{\prime2})-36\lambda g^{\prime2}$\tabularnewline
$\gamma_{c_{T}}$  & $3g^{\prime4}+\frac{9}{8}g^{\prime2}g^{2}+3\lambda g^{\prime2}$  & $\frac{9}{8}g^{\prime2}g^{2}$  & $-\frac{9}{4}g^{\prime2}g^{2}-6\lambda g^{\prime2}$ & $-\frac{9}{4}g^{\prime2}g^{2}$\tabularnewline
$\gamma_{c_{2B}}$  & $\frac{94}{3}g^{\prime2}$ & $0$  & $-\frac{2}{3}g^{\prime2}$ & $0$ \tabularnewline
$\gamma_{c_{2W}}$  & $\frac{53}{12}g^{\prime2}\left(1-3t_{\theta_{W}}^{2}\right)$ & $\frac{331}{12}g^{2}+\frac{29}{4}g^{\prime2}$  & $0$ & $-\frac{2}{3}g^{2}$ \tabularnewline
$\gamma_{c_{B}}$  & \textbf{$\frac{59}{4}g^{\prime2}$} & $-\frac{g^{2}}{4}$  & $\frac{g^{\prime2}}{6}+6y_{t}^{2}$ & $\frac{g^{2}}{2}$\tabularnewline
$\gamma_{c_{W}}$  & $\left(\frac{29}{8}-\frac{53}{4}t_{\theta_{W}}^{2}\right)g^{\prime2}$  & $\frac{79}{8}g^{2}+\frac{29}{4}g^{\prime2}$  & $\frac{g^{\prime2}}{6}$ & $\frac{17}{2}g^{2}+6y_{t}^{2}$\tabularnewline
others  & $0$ & $0$ & $0$ & $0$ \tabularnewline
\end{tabular}

\vspace{0.5cm}

\begin{tabular}{C{0.88cm} | C{3.115cm} C{3.115cm} C{3.115cm} C{3.115cm} }
 & $c_{BB}$  & $c_{HB}$ & $c_{HW}$ & $c_{3W}$ \tabularnewline
\hline 
$\gamma_{c_{H}}$  & $0$ & $-\frac{9}{4}g^{\prime2}(g^{\prime2}-2g^{2})-6\lambda g^{\prime2}$ & $-\frac{9}{4}g^{2}(g^{2}-2g^{\prime2})-36\lambda g^{\prime2}$& $0$ \tabularnewline
$\gamma_{c_{T}}$  & $0$ & $-\frac{9}{4}g^{\prime2}g^{2}-6\lambda g^{\prime2}$ & $-\frac{9}{4}g^{\prime2}g^{2}$& $0$ \tabularnewline
$\gamma_{c_{2B}}$  & $0$ & $-\frac{2}{3}g^{\prime2}$ & $0$ & $0$ \tabularnewline
$\gamma_{c_{2W}}$  & $0$ & $0$ & $-\frac{2}{3}g^2$& $0$   \tabularnewline
$\gamma_{c_{B}}$  & $8g^{\prime 2}$  & $-\frac{9}{2}g^2-\frac{1}{3}g^{\prime 2}-4\lambda$ & $-\frac{17}{2}g^2+8\lambda$  & $-12g^2$ \tabularnewline
$\gamma_{c_{W}}$ &  $0$   & $-\frac{5}{6}g^{\prime2}$  & $11g^2+\frac{1}{2}g^{\prime 2}-12\lambda$  & $10g^2$ \tabularnewline
$\gamma_{c_{BB}}$  & $-\frac{3 }{2}g^{\prime2}-\frac{9}{2}g^2+6y_{t}^{2}+12\lambda$  & $\frac{3}{2}g^2-2\lambda$  & $\frac{3}{2}g^{2}-2\lambda$  & $3g^{2}$\tabularnewline
$\gamma_{c_{HB}}$  & $-8g^{\prime 2}$  & $\frac{1}{2}g^{\prime2}+\frac{9}{2}g^2+6y_{t}^{2}+4\lambda$  & $9g^{2}-8\lambda$  & $12g^2$ \tabularnewline
$\gamma_{c_{HW}}$  & $0$  & $g^{\prime 2}$  & $-\frac{5}{2}g^2-\frac{1}{2}g^{\prime 2}+6y_t^2+12\lambda$  & $-10g^2$ \tabularnewline
$\gamma_{c_{3W}}$  & $0$ & $0$ & $0$  & $\frac{53}{3} g^2$ \tabularnewline
\end{tabular}
\caption{\small Anomalous dimension matrix for the Wilson coefficients of the dim-6 bosonic operators, in the SILH basis~\cite{Giudice:2007fh}.\label{tab:OurBasisMatrix3}}
\end{center}
\end{table}
}



\newpage
\begin{thebibliography}{99}


\bibitem{Higgs1}
  G.~Aad {\it et al.}  [ATLAS Collaboration],
  Phys.\ Lett.\ B {\bf 716} (2012) 1
  [arXiv:1207.7214 [hep-ex]].
\bibitem{Higgs2}
  S.~Chatrchyan {\it et al.}  [CMS Collaboration],
  Phys.\ Lett.\ B {\bf 716} (2012) 30
  [arXiv:1207.7235 [hep-ex]].

  
  \bibitem{Buchmuller:1985jz}
  W.~Buchm\"uller and D.~Wyler,
  Nucl.\ Phys.\ B {\bf 268} (1986) 621.

\bibitem{Narison:1983}
  S. Narison and R. Tarrach, Phys. \ Lett. \ {\bf B125} (1983) 217.
  
\bibitem{Morozov:1985ef}
  A.~Y.~Morozov,
  Sov.\ J.\ Nucl.\ Phys.\  {\bf 40} (1984) 505
   [Yad.\ Fiz.\  {\bf 40} (1984) 788].

\bibitem{Hagiwara:1993ck}
  K.~Hagiwara, S.~Ishihara, R.~Szalapski and D.~Zeppenfeld,
  Phys.\ Rev.\ D {\bf 48} (1993) 2182.

  
\bibitem{Hagiwara:1994pw}
  K.~Hagiwara, S.~Matsumoto, D.~Haidt and C.~S.~Kim,
  Z.\ Phys.\ C {\bf 64} (1994) 559
   [Erratum-ibid.\ C {\bf 68} (1995) 352]
  [\arXhref{hep-ph/9409380}].

\bibitem{Alam:1997nk}
  S.~Alam, S.~Dawson and R.~Szalapski,
  Phys.\ Rev.\ D {\bf 57} (1998) 1577
  [\arXhref{hep-ph/9706542}].

\bibitem{Barbieri:2007bh}
  R.~Barbieri, B.~Bellazzini, V.~S.~Rychkov and A.~Varagnolo,
  Phys.\ Rev.\ D {\bf 76} (2007) 115008
  [arXiv:\arXhref{0706.0432} [hep-ph]].
  
\bibitem{Grojean:2013kd}
  C.~Grojean, E.~E.~Jenkins, A.~V.~Manohar and M.~Trott,
  JHEP {\bf 1304} (2013) 016
  [arXiv:\arXhref{1301.2588} [hep-ph]].

\bibitem{Elias-Miro:2013gya}
  J.~Elias-Miro, J.~R.~Espinosa, E.~Masso and A.~Pomarol,
  JHEP {\bf 1308} (2013) 033
  [arXiv:\arXhref{1302.5661} [hep-ph]].

\bibitem{EEMP2}
  J.~Elias-Miro, J.~R.~Espinosa, E.~Masso and A.~Pomarol,
  JHEP {\bf 1311} (2013) 066
  [arXiv:\arXhref{1308.1879} [hep-ph]].
  

\bibitem{Jenkins:2013zja}
  E.~E.~Jenkins, A.~V.~Manohar and M.~Trott,
  JHEP {\bf 1310} (2013) 087
  [arXiv:\arXhref{1308.2627} [hep-ph]].
  
\bibitem{Jenkins:2013wua}
  E.~E.~Jenkins, A.~V.~Manohar and M.~Trott,
  [arXiv:\arXhref{1310.4838} [hep-ph]].
    
\bibitem{Grzadkowski:2010es}
  B.~Grzadkowski, M.~Iskrzynski, M.~Misiak and J.~Rosiek,
  JHEP {\bf 1010} (2010) 085
  [arXiv:\arXhref{1008.4884} [hep-ph]].

\bibitem{Contino:2013kra}
  R.~Contino, M.~Ghezzi, C.~Grojean, M.~Muhlleitner and M.~Spira,
  JHEP {\bf 1307} (2013) 035
  [arXiv:\arXhref{1303.3876} [hep-ph]].

\bibitem{Dumont:2013wma}
  B.~Dumont, S.~Fichet and G.~von Gersdorff,
  JHEP {\bf 1307} (2013) 065
  [arXiv:\arXhref{1304.3369} [hep-ph]].

\bibitem{Pomarol:2013zra} 
  A.~Pomarol and F.~Riva,
  [arXiv:\arXhref{1308.2803} [hep-ph]].

  
\bibitem{DeRujula:1991se}
  A.~De Rujula, M.~B.~Gavela, P.~Hernandez and E.~Masso,
  Nucl.\ Phys.\ B {\bf 384} (1992) 3.

\bibitem{Mebane:2013zga}
  H.~Mebane, N.~Greiner, C.~Zhang and S.~Willenbrock,
  Phys.\ Rev.\ D {\bf 88} (2013) 015028
  [arXiv:\arXhref{1306.3380} [hep-ph]].

\bibitem{Chen:2013kfa}
  C.~-Y.~Chen, S.~Dawson and C.~Zhang,
  [arXiv:\arXhref{1311.3107} [hep-ph].
 
   \bibitem{Barbieri} 
  R.~Barbieri, A.~Pomarol, R.~Rattazzi and A.~Strumia,
  Nucl.\ Phys.\ B {\bf 703}, 127 (2004)
  [\arXhref{hep-ph/0405040}].
 
\bibitem{Han:2004az}
  Z.~Han and W.~Skiba,
  Phys.\ Rev.\ D {\bf 71} (2005) 075009
  [\arXhref{hep-ph/0412166}].

\bibitem{Cacciapaglia:2006pk}
  G.~Cacciapaglia, C.~Csaki, G.~Marandella and A.~Strumia,
  Phys.\ Rev.\ D {\bf 74} (2006) 033011
  [\arXhref{hep-ph/0604111}].

\bibitem{Grojean:2013nya}
  C.~Grojean, E.~Salvioni, M.~Schlaffer and A.~Weiler,
  arXiv:1312.3317 [hep-ph].

 \bibitem{Grinstein:1991cd}
  B.~Grinstein and M.~B.~Wise,
  Phys.\ Lett.\ B {\bf 265} (1991) 326.
 
 \bibitem{leptgc} 
  S.~Schael {\it et al.}  [ALEPH and DELPHI and L3 and OPAL and LEP Electroweak Collaborations],
  Phys.\ Rept.\  {\bf 532}, 119 (2013)
  [arXiv:\arXhref{1302.3415} [hep-ex]].
 
 \bibitem{Giudice:2007fh}
  G.~F.~Giudice, C.~Grojean, A.~Pomarol and R.~Rattazzi,
  JHEP {\bf 0706} (2007) 045
  \mbox{[\arXhref{hep-ph/0703164}]}.

\bibitem{Barbieri:1987fn}
  R.~Barbieri and G.~F.~Giudice,
  Nucl.\ Phys.\ B {\bf 306} (1988) 63.
  
\bibitem{Gounaris:1996rz}
  G.~Gounaris, J.~L.~Kneur, D.~Zeppenfeld, Z.~Ajaltouni, A.~Arhrib, G.~Bella, F.~A.~Berends and M.~S.~Bilenky {\it et al.},
  \arXhref{hep-ph/9601233}.

\bibitem{gfitter} 
  M.~Baak, M.~Goebel, J.~Haller, A.~Hoecker, D.~Kennedy, R.~Kogler, K.~Moenig and M.~Schott {\it et al.},
  Eur.\ Phys.\ J.\ C {\bf 72}, 2205 (2012)
  [arXiv:\arXhref{1209.2716} [hep-ph]].

\bibitem{Francesco} 
  We thank F. Riva for providing us with the latest constraint on $c_H$ from the fit in ref.~\cite{Pomarol:2013zra}.

\bibitem{Ciuchini:2013pca}
  M.~Ciuchini, E.~Franco, S.~Mishima and L.~Silvestrini,
  JHEP {\bf 1308} (2013) 106
  [arXiv:\arXhref{1306.4644} [hep-ph]].
  
\bibitem{Grojean:2013qca}
  C.~Grojean, O.~Matsedonskyi and G.~Panico,
  JHEP {\bf 1310} (2013) 160
  [arXiv:\arXhref{1306.4655} [hep-ph]].
  
\bibitem{Contino:2013gna}
  R.~Contino, C.~Grojean, D.~Pappadopulo, R.~Rattazzi and A.~Thamm,
  [arXiv:\arXhref{1309.7038} [hep-ph]].

\bibitem{Dawson:2013bba} 
  S.~Dawson, A.~Gritsan, H.~Logan, J.~Qian, C.~Tully, R.~Van Kooten, A.~Ajaib and A.~Anastassov {\it et al.},
  [arXiv:\arXhref{1310.8361} [hep-ex]].

\bibitem{ILC} 
  H.~Baer, T.~Barklow, K.~Fujii, Y.~Gao, A.~Hoang, S.~Kanemura, J.~List and H.~E.~Logan {\it et al.},
  [arXiv:\arXhref{1306.6352} [hep-ph]].

\bibitem{Gomez-Ceballos:2013zzn}
  M.~Bicer, H.~Duran Yildiz, I.~Yildiz, G.~Coignet, M.~Delmastro, T.~Alexopoulos, C.~Grojean and S.~Antusch {\it et al.},
  [arXiv:\arXhref{1308.6176}~[hep-ex]].
    
\bibitem{Domenech:2012ai}
  O.~Domenech, A.~Pomarol and J.~Serra,
  Phys.\ Rev.\ D {\bf 85} (2012) 074030
  [arXiv:\arXhref{1201.6510} [hep-ph]].
 J.~Serra, talk at Planck 2012 [\href{http://planck12.fuw.edu.pl/talks/serra.pdf}{http://planck12.fuw.edu.pl/talks/serra.pdf}].

\bibitem{Simmons:1995hb}
  E.~H.~Simmons and P.~L.~Cho,
  In *Los Angeles 1995, Vector boson self-interactions* 323-334
  [\arXhref{hep-ph/9504401}].

\bibitem{AJMT}
  R.~Alonso, E.~Jenkins, A. Manohar and M.~Trott,
  [arXiv:\arXhref{1312.2014} [hep-ph]].
 
\bibitem{Pineda:2001ra}
  A.~Pineda,
  Phys.\ Rev.\ D {\bf 65} (2002) 074007
  [\arXhref{hep-ph/0109117}].


\end{thebibliography}
\end{document}